\documentclass[11pt,graphicx,subfigure,axodraw]{article}
\usepackage{multirow}
\usepackage{amsfonts}
\usepackage{amssymb}
\usepackage{epsfig}
\usepackage[usenames,dvipsnames]{color}
\newcommand{\beqn}{\begin{eqnarray}}
\newcommand{\eeqn}{\end{eqnarray}}
\newcommand{\be}{\begin{equation}}
\newcommand{\ee}{\end{equation}}
\newcommand{\non}{\nonumber \\}

\def\s1{$s_{\alpha}$}
\def\s2{$s_{\gamma}$}
\def\s3{$s_{\delta}$}
\def\c1{$c_{\alpha}$}
\def\c2{$c_{\gamma}$}
\def\c3{$c_{\delta}$}

\def\45{\overline{45}}
\def\5{\overline{5}}
\def\70{\overline{70}}
\def\50{\overline{50}}
\def\a{/anti-triplets~}
\newcommand{\ov}{\overline }

\begin{document}
\baselineskip 18pt
\begin{titlepage}

\begin{flushright}
NUB-TH:3267\\
OSU-HEP-11-09\\
\end{flushright}

\begin{center}
{\bf {\ {\textsf{\Large{
 {\boldmath
 Variety of
 $SO(10)$ GUTs with Natural Doublet-Triplet  Splitting via the Missing Partner  Mechanism  }
    }}}}}

\vskip 0.5 true cm \vspace{0cm}
\renewcommand{\thefootnote}
{\fnsymbol{footnote}}
 K.S. Babu$^a$, Ilia Gogoladze$^b$, Pran Nath$^c$ and Raza M. Syed$^{c,d}$
\vskip 0.5 true cm
\end{center}
\begin{center}
\noindent {$^a$\textit{Department of Physics, Oklahoma State University, Stillwater, OK, 74078, USA} \\
$^b$\textit{Bartol Research Institute, Department of Physics and Astronomy,
University of Delaware, Newark, DE 19716, USA} \\
$^c$\textit{Department of Physics, Northeastern University,
Boston, MA 02115-5000, USA} \\
$^d$\textit{Department of Physics, American University of Sharjah,
P.O. Box 26666,
Sharjah, UAE\footnote{Permanent address}}
} \\
\end{center}
\date{\today}
\vskip 1.0 true cm \centerline{\bf Abstract}

We  present a new class of unified $SO(10)$ models where the GUT symmetry breaking down to the
standard model gauge group involves just one scale, in contrast to the conventional $SO(10)$ models
which require two scales.  Further, the models we discuss possess a natural
doublet-triplet splitting via the missing partner mechanism without fine tuning.  Such models involve
$560+\ov{560}$ pair of heavy Higgs fields along with a set of light fields. The $560+\ov{560}$
are the simplest representations of $SO(10)$ besides the $126+\ov{126}$ which contain an excess of color triplets
over $SU(2)_L$ doublets.  We discuss several possibilities for realizing the missing partner mechanism
within these schemes.  With the $126+\ov{126}$ multiplets, three viable models are found with additional
fields belonging to $\{210 + 2 \times 10 + 120\}$,   $\{ 45 + 10 + 120\}$, or $\{210 + 16 + \ov{16} + 10 + 120\}$.
With the $560+\ov{560}$, a unique possibility arises for the missing partner mechanism, with additional  $\{2\times 10+ 320\}$
fields.  These models are developed in some detail. It is shown that fully realistic fermion masses can arise in some  cases, 
while others can be made realistic by addition of vector--like representations.  Naturally large neutrino mixing angles, 
including sizable $\theta_{13}$, can emerge in these models.  The couplings of the $H_u(H_d)$  Higgs doublets of the MSSM 
which give masses to the up quarks (down quarks and leptons) are not necessarily equal at the grand unification scale and 
would lead to a new phenomenology at the low energy scales.

\medskip
\noindent

\end{titlepage}
\section{Introduction}
Gauge symmetry based on $SO(10)$ provides a framework for unifying the $SU(3)_C\times SU(2)_L\times U(1)_Y$ gauge
groups and for unifying quarks and leptons in a single $16$--plet spinor representation \cite{georgi}.
Additionally, the 16--plet also contains a right--handed
singlet state, which is needed to give mass to the neutrino via the seesaw mechanism. Supersymmetric
$SO(10)$ models have the added attraction that they predict correctly the unification of gauge couplings, and
solve the hierarchy problem by virtue of SUSY.  However, SUSY $SO(10)$ models, as usually constructed, have
two drawbacks, both related to the symmetry breaking sector.  
 Typically three types of Higgs fields are needed, e.g., $16+\ov {16}$ for rank reduction,
$45$ for breaking the symmetry down to the standard model symmetry, and a $10$ for electroweak symmetry breaking.
The above implies that two different mass scales  are involved in breaking of the GUT symmetry, one to
reduce the rank and the other to reduce the symmetry all the way to $SU(3)_C\times SU(2)_L\times U(1)_Y$.
Second, one must do an extreme fine--tuning at the level of one part in $10^{14}$ to get the Higgs doublets
of MSSM light, while rendering super--heavy masses to their color--triplet GUT partners.
We have previously investigated $SO(10)$ unified models wherein the first problem is addressed
with some success. A single pair of $144+\overline{144}$ dimensional vector-spinor Higgs multiplets  can break the $SO(10)$ gauge symmetry in a single  step all the way down to Standard Model (SM) gauge symmetry $SU(3)_C \times SU(2)_L \times U(1)_Y$ \cite{Babu:2005gx,Babu:2006rp,Nath:2005bx}.
There we also exhibited the possibility to obtain from the same irreducible $144+\overline{144}$
multiplets a pair of light Higgs doublets, necessary for the breaking of the
electroweak symmetry. However, the second problem mentioned above, of extreme fine--tuning for making the
Higgs doublets light, was not solved in that framework.

This doublet-triplet (DT) splitting problem is quite generic in grand
unified models. One remedy proposed to solve the problem is the so called missing VEV mechanism \cite{DW}
where the vacuum expectation value (VEV) of a $45$ Higgs field which breaks the $SO(10)$ symmetry lies in the $(B-L)$--preserving
direction, and generates masses for the Higgs triplets but not for the Higgs doublets from a $10$--plet.
This mechanism works in $SO(10)$ and  has no analog in $SU(5)$.
A second remedy is the missing partner mechanism, first discovered
in the context of $SU(5)$ \cite{SU(5)Missingpartner}
and later investigated for $SO(10)$ \cite{SO(10)Missingpartner}.
In $SU(5)$, the missing partner mechanism requires
the representations $50+\ov{50}+ 75$ (all heavy)
in addition to the light $5+\bar 5$ of Higgs.\footnote{Heavy in this context means the presence of a GUT scale mass term for the field, while
light means the absence of a GUT scale mass for the field. Components of a light field can become heavy via its mixing with a
heavy field.} Now, it turns out the $50+\ov{50}$ have one pair of Higgs triplets\a and  no Higgs doublets. Thus when they mix with the $5+\bar 5$,
the Higgs triplets\a in $5+\bar 5$ become heavy,
while the Higgs doublets in $5+\bar 5$ remain light since there are no Higgs
doublets in $50+\ov{50}$ for them to pair with.  The $75$--plet Higgs is needed to induce the mixing of the color triplets from the $5+\bar 5$
and the $50+\bar 50$.  In Ref. \cite{SO(10)Missingpartner} the missing partner mechanism was extended to
$SO(10)$ by considering the set of heavy fields $126+\ov{126}+210$ and a set of light fields.  The lowest dimensional representations
of $SO(10)$ that contain  $50+\ov{50}$ of $SU(5)$ are the $126+\ov{126}$.  The $210$ is needed since it contains the $75$ of $SU(5)$.

In this paper we first investigate systematically the missing partner $SO(10)$ models anchored by the $126+\ov{126}$.
Here two new possibilities are found, in addition to the model proposed in Ref. \cite{SO(10)Missingpartner}.
Then we discuss a new
class of $SO(10)$ models, where we solve the DT splitting problem via the missing partner mechanism, and simultaneously achieve
the one-step GUT symmetry breaking of $SO(10)$.  Here the missing partner mechanism is anchored by a pair of $560+\ov{560}$ Higgs fields, which
also contain $50+\ov{50}$ under $SU(5)$.  These fields are the next simplest representations containing an excess of color triplets over
$SU(2)_L$ doublets beyond $126+\ov{126}$. In addition, the $560$ contains in the $SU(5)\times U(1)$ decomposition $1(-5)+24(-5)+75(-5)$. When
these fields acquire vacuum expectation values (VEVs), one ends up with the residual gauge
group $SU(3)_C\times SU(2)_L\times U(1)_Y$.  Thus $SO(10)$ gauge symmetry breaks down to the Standard Model in one step.
An attractive feature of the $560+\ov{560}$ combination is that they contain $75+\ov{75}$--plet representations which is absent in the $126+\ov{126}$ pair, necessitating the inclusion of the $210$--plet in that case.
($75$--plet of $SU(5)$ is self-dual, we use the notation $75$ to denote $75(-5)$ and $\ov{75}$ to denote a $75(+5)$ with opposite
$U(1)$ quantum numbers under $SO(10) \rightarrow SU(5) \times U(1)$.)
The missing partner mechanism works in the case of $560$ by its mixing with certain light fields.  A unique possibility is found, with
the light fields belonging to $\{2\times 10+320\}$ representations.  A non-trivial constraint on such models
for consistency is that all exotics (particle other than the Higgs doublets of the MSSM) from
the $560+\ov{560}$ or $126+\ov{126}$ must acquire GUT scale masses, which is satisfied in the models presented here.

Some of the models developed lead to fully realistic fermion masses and mixings.  For those which do not,
we suggest a method of adding equal number of heavy and light fields to the Higgs spectrum outlined above, which leaves the
missing partner mechanism intact.  The MSSM fields $H_u$ and $H_d$ will now have an admixture of these newly added light
field, which in turn makes the fermion masses fully realistic.  In particular, we find that these models can naturally lead to large
neutrino mixing angles, including a sizable $\theta_{13}$.

 The outline of the rest of the paper is as follows: In Sec. 2 we describe the essence of the missing partner
 mechanism in $SU(5)$. In Sec. 3 we discuss the missing partner mechanism anchored by $126+\ov{126}$.  Here we
 identify three viable models which require no fine tuning.
 In Sec. 4 we discuss the newly proposed missing partner mechanism anchored by $560+\ov{560}$ Higgs fields.
 Since the models based on $560+\ov{560}$ are
 new and not familiar we discuss further details of these in Secs. 5-8.  Thus in Sec. 5 we discuss the
 560--plet as a constrained spinor-tensor multiplet,
 and exhibit explicitly its components  for calculational purposes. In Sec. 6 we discuss the breaking of the GUT
 symmetry by $560+\ov{560}$ multiplet so that $SO(10)\to SU(3)_C\times SU(2)_L\times U(1)_Y$
 using a single  mass scale.  Numerical solutions are presented to the six coupled minimization conditions
involving the VEVs of the sub-multiplets $1, 24, 75$ and $\bar 1, \overline{24},\overline{75}$.
In Sec. 7 we compute the Higgs doublet and the Higgs triplet mass  matrices in one
model, i.e., the model with the Higgs fields $560+\ov{560} + (2\times 10 + 320)$.
A brief discussion of how realistic fermion masses can arise in these models is presented in Sec. 8.
Conclusions are given in Sec. 9.  Some further calculational details
on spontaneous breaking with $560+\ov{560}$ multiplet are given in the Appendix.

\section{Missing Partner Mechanism in $SU(5)$}

Before discussing the missing partner mechanism in $SO(10)$ we briefly review the simpler case of missing
partner mechanism in  $SU(5)$.  The simplest Higgs structure which breaks $SU(5)$ and contains
Higgs doublets needed for the breaking of the electroweak symmetry consists of $5(H_2)+\bar 5(H_1)$
and a $24$--plet of Higgs with a superpotential of the type
\beqn
  \lambda \left [\frac{1}{2} M Tr(\Sigma^2) +  \frac{1}{3} Tr(\Sigma^3) \right]
   + \lambda'  H_{1x} (\Sigma^x_y + 2M' \delta^x_y)H_2^y.
   \label{su5}
\eeqn
Here, if the 24-plet develops  a VEV  of the form $\left \langle \Sigma \right\rangle= V\,{\rm diag}\,(2,2,2, -3, -3)$ arising from the minimization
of the first  two terms  in Eq. (\ref{su5}), then  $SU(5)$ breaks  down to $SU(3)_C\times SU(2)_L\times
U(1)$. However, the Higgs doublets and Higgs triplets\a in $5+\bar 5$ would gain super--heavy masses and
a fine tuning of one part in $10^{14}$ is needed to make the Higgs doublets light, without also making
the color triplets light.

To obtain  Higgs doublets naturally light without fine tuning, one can utilize a different array of Higgs multiplets
which are $5(H_2^i), \bar 5 (H_{1j}),50^{ijk}_{lm}, \overline{50}^{ij}_{klm}, 75^{ij}_{kl}$,
where the fields satisfy the following constraints: $50^{ijk}_{lk}=0, \, 50^{ijk}_{jk}=0, \, \overline{50}_{lmn}^{in}=0,\,
\overline{50}_{lmn}^{mn}=0,\, 75^{ij}_{kj}=0,\,75^{ij}_{ij}=0$. We consider now a superpotential of the form
\beqn
W_{Higgs}= W_0(75) + M \,50\cdot\overline{50} + \lambda_1 50\cdot75\cdot\bar 5 + \lambda_2 \overline{50}\cdot75\cdot5.
\label{su5higgs}
\eeqn
Here $W_0(75) = M_{75}\, 75^2 + \lambda \,75^3$, which generates a VEV for the Standard Model singlet component of $75$
and breaks the $SU(5)$ symmetry down to $SU(3)_C \times SU(2)_L \times U(1)_Y$.
In the $SU(3)_C\times SU(2)_L\times U(1)$ decomposition the multiplets break as follows:
\beqn
5 &= &(1,2)(3) + (3,1)(-2),\nonumber\\
50&=&(1,1)(-12) + (3,1)(-2) +  (\bar 3, 2) (-7) + (\bar 6,3) (-2)
+(6,1)(8) + (8,2)(3),\nonumber\\
75&=&(1,1)(0) + (3,1)(10) + (3,2) (-5) + (\bar 3, 1) (-10) + (\bar 3, 2)(5) +\nonumber\\
& & (\bar 6,2)(-5)  + (6,2)(5) + (8,1)(0) + (8,3)(0)\,,
\label{5075}
\eeqn
with the decomposition for the $\bar 5$ and $\ov{50}$  obvious from Eq. (\ref{5075}).
(The SM hypercharge $Y/2$ is $1/6$ times the charge quoted above.)
The decomposition above shows that $5+\bar 5$ have one pair of Higgs doublets and one pair of
Higgs triplets/anti-triplets while the $50+\overline{50}$ have one pair of  Higgs triplets/anti-triplets
and no Higgs doublets. Thus after the GUT symmetry breaks
via the VEV of the $75$--plet, the last three terms in the superpotential of Eq. (\ref{su5higgs}) would
generate masses for the Higgs triplets/anti-triplets while the Higgs doublets remain light because there are
no Higgs doublets to pair up with in $50+ \ov {50}$. So this is the simplest way the missing partner
mechanism works in $SU(5)$.
Of course, as a GUT group $SO(10)$ is more desirable than
$SU(5)$ and thus it is of interest to discuss the various possibilities to achieve the missing partner
mechanism in $SO(10)$. The main aim of this paper is to explore these possibilities.

\section{Missing Partner Mechanism in $SO(10)$ Anchored by $126+\ov{126}$}

 In searching for the missing partner mechanism in $SO(10)$ we note once again
 that the mechanism works in $SU(5)$
 because  of the central fact that the $50+\ov{50}$ multiplets in $SU(5)$
 contain an $SU(3)_C$ triplet/anti-triplet pair but no $SU(2)_L$ doublets.
 Thus to implement the mechanism at the level of $SO(10)$, a simple procedure would be to
 look for $SO(10)$ representations which contain the $SU(5)$ representations $50+\ov{50}$.  Now the
 lowest irreducible representation  of $SO(10)$ where the $\ov{50}(50)$ appears is
the $126(\ov{126})$ while the next one is  $560(\ov{560})$ \cite{slansky}.  The case $126+\ov{126}$ has
been studied in  \cite{SO(10)Missingpartner}.  Here we systematically investigate this
case, and find three viable solutions without fine tuning, two of which are new.
The $126$ decomposes  under $SU(5)\times U(1)_X$ as follows:
\beqn
126= 1(-10)+ \bar 5(-2)+ 10(-6) + \ov{15}(6)+ 45(2) + \ov{50}(-2).
\eeqn
Notice that the 126--plet does not contain the 75--plet of $SU(5)$, and thus to break the GUT symmetry we include
a 210 representation which contains the 75--plet as is seen from the following  decomposition
under $SU(5)\times U(1)$,
\beqn
210=1(0) + 5(-8) + \bar 5(8) + 10(4) + \ov{10}(-4)+ 24(0) + 40(-4) +\ov{40} (4) + 75(0).
\eeqn
 Thus to achieve DT splitting in this case
one needs a heavy sector consisting of  $(126+\ov{126}+210)$.  Since $126+\ov{126}$
contain 2 doublet pairs and 3 triplet\a pairs, and since 210 contains
one doublet pair and one triplet\a pair,  one has a total of
(3,4) (doublet, triplet\a) pairs from the heavy sector. Suppose the light sector
contains  (4,4) (doublet, triplet\a) pairs.  If the light sector obtains mass only by mixing with
the heavy sector, one will be left with just one pair of light Higgs doublets while
all the triplets\a will be heavy.  We now discuss various ways of achieving this.

\vspace*{0.1in}
 \noindent{\bf (i) Heavy  \{\mbox{\boldmath{$126+\ov{126} +  210$}}\} + Light  \{\mbox{\boldmath{$2 \times 10 +  120$}}\} model:}
\vspace*{0.1in}

 Now, suppose we choose a set of $\{2 \times 10 + 1 \times 120\}$ light fields, along with $\{126+\ov{126}+210\}$ heavy fields.
 These light fields will have (4,4) (doublet, triplet\a), as can be seen from the following decompositions
 under $SU(5) \times U(1)$:
 \begin{eqnarray}
 10&=&5(2)+\ov{5}(-2) \nonumber \\
120&=&5(2) + \ov{5}(-2)  + 10(-6) + \ov{10}(6)+ 45(2) +\ov{45}(-2).
\label{120}
\end{eqnarray}
Each of the $10$'s contains one (doublet, triplet\a) pairs, while the $120$ contains two such pairs,
one pair from the $5+\ov{5}$ and one pair from the $45+\ov{45}$ fragments of Eq. (\ref{120}).  (The
$45$ of $SU(5)$ decomposes under $SU(3)_C \times SU(2)_L \times U(1)$ as $45 = (1,2)(3)+(3,1)(-2)+(3,3)(-2)+(\ov{3},1)(8)+
(\ov{3},2)(-7)+(\ov{6},1)(-2)+(8,2)(3)$, which shows its doublet/triplet content.)
These light fields mix with the heavy $\{126+\ov{126}+210\}$ fields through the
following set of couplings:
  \beqn \label{210}
W_{DT}^{(i)} =    10_i\cdot126\cdot 210 + 10_i\cdot \ov{126}\cdot 210 + 120\cdot126\cdot210 + 120\cdot\ov{126}\cdot 210~.
\label{W1}
\eeqn
Notice the absence of bare mass terms (or effective mass terms via the couplings $120^2 \cdot 210$ and
$120\cdot 10_i \cdot 210$) for the light fields in $\{2 \times 10+120\}$.  As per the counting
listed above, these couplings would lead to 4 pairs of super--heavy color triplet/anti-triplets, and
3 pairs of super--heavy doublets, leaving one pair of light Higgs doublets, to be identified with $H_u$ and $H_d$
of MSSM. Further, all the remaining components of the light fields which do not enter in the DT splitting
gain super--heavy masses.  This point is quite non--trivial for the sub-multiplets from $120$,
but all of these fields pair up with sub-multiplets from the $126+\ov{126}$ or $210$, as we now show.

In the $SU(5)\times U(1)$ decomposition the exotics in 120
are $10(-6), \ov{10}(6), 45(2),  \ov{45}(-2)$, see Eq. (\ref{120}).
The relevant superpotential that achieves natural DT splitting is given in Eq. (\ref{W1}).
After spontaneous breaking the $75(0)$--plet in 210 develops a VEV and generates
mass for the exotics as follows.
\beqn
\left\langle({210},75(0))\right\rangle \cdot({126}, 10(-6))\cdot (120, (\ov{10}(6)),
~~~~\langle ({210},75(0)) \rangle \cdot(\ov{126}, \ov{10}(6))\cdot (120, 10(-6)),\nonumber\\
\langle ({210},75(0)) \rangle \cdot({126}, 45(2))\cdot (120, \ov{45}(-2)),
~~~~\langle ({210},75(0)) \rangle \cdot(\ov{126}, \ov{45}(-2))\cdot (120, 45(2))~.
\label{exoticmass}
\eeqn
Here the sub-multiplets under $SU(5) \times U(1)$ involved are explicitly indicated.  We see that the $75(0)$ VEV
of the 210--plet alone would give super-heavy masses to all exotics from the 120.
There are additional contributions to the exotic masses from the VEVs of $24(0)$ and $1(0)$ fragments of the $210$,
which are analogous to the ones in Eq. (\ref{exoticmass}).
When the singlet VEVs $1(-10)$ from the $126$ and $(1,+10)$ from the $\overline{126}$ are inserted in Eq. (\ref{W1}), additional mass terms
would arise which mix the $\ov{10}(4)$ from the 120 with the $10(6)$ from the 210, and similarly the $10(-4)$ from the 120
with the $\ov{10}(-6)$ from the 210.   Note that
one combination of $10+\overline{10}$ pair under $SU(5)$ from the $210+126+\overline{126}$ are the Goldstone
modes, but the exotic $10+\overline{10}$ components of $120$ pair up with the orthogonal non--Goldstone components
from the $210+126+\overline{126}$.  Thus all the exotics in $120$--plet are removed from the light spectrum.

The model just described is fully realistic.  Symmetry breaking occurs consistently \cite{aulakh}, there
are no unwanted light states, and the masses and mixings of quarks and leptons are induced correctly.
The last fact follows from the two flavor symmetric sets of Yukawa couplings with the $10$--plets of Higgs
which contribute equally to down quark and charged lepton masses,
an antisymmetric set of Yukawa coupling with the 120--plet which distinguishes down quarks and charged
leptons, and the Yukawa couplings to the $\ov{126}$--plet which generates heavy masses for the right--handed
neutrinos.  In contrast, when the missing partner mechanism is employed in $SU(5)$, the degeneracy between
the down quarks and charged leptons is not lifted.  The simplest way out of the wrong mass predictions in
this case would be to also add a vector--like pair of fermions in the $10+\ov{10}$ representations, the
components of which mix by different amounts with the down quarks and the charged leptons through couplings such as
$10_i \ov{10}\,\,75$ ($i=1-3$ is the family index). Alternatively, one could resort to higher dimensional
non--renormalizable operators.  Generating neutrino masses via the seesaw mechanism
would also require the introduction of a singlet fermion sector in $SU(5)$.

What about other possibilities for the light states in missing partner $SO(10)$ models with
$\{126+\ov{126}+210\}$ heavy fields?  Since the heavy sector has four pairs of color triplet/anti--triplet fields,
it might appear that mixing four 10--plets with these
heavy fields could also lead to the desired light Higgs doublets.  However, this is not so.  In this case,
only two combinations of the four 10--plets will get involved in the light--heavy mixings analogous to Eq. (\ref{210}),
resulting in two 10--plets entirely becoming light.  Similarly, adding pairs of $16+\ov{16}$ which contain one
pair of Higgs doublets and a pair of color triplet/anti--triplet does not work in any simple way.  To emphasize this
point, let us consider replacing some of the doublets and color triplets/anti-triplets of the light sector in the
model described above by those from $16+\ov{16}$.
Under $SO(10) \rightarrow SU(5) \times U(1)$ we have $16 = 1(-5)+\ov{5}(3)+10(-1)$.  Thus $16+\ov{16}$ contain one
pair of doublets and one pair of color triplet/anti-triplets.  They also contain an exotic $10+\ov{10}$ pair under $SU(5)$ subgroup,
which must be given large mass.  Now, the heavy sector $\{126+\ov{126}+210\}$ contains two pairs of $10+\ov{10}$, one of which
pairs up with the gauge super--multiplet upon symmetry breaking.  Thus, at most one $10 + \ov{10}$ exotic from the
$16+\ov{16}$ can be paired with those from the heavy fields.  This implies that in the example with $\{2 \times 10+
1 \times 120\}$ light fields, we cannot replace the $120$ field -- which contains two pairs of doublets and triplets --
by 2 pairs of $16+\ov{16}$.  Replacing one light $10$--plet by a $16+\ov{16}$ might appear feasible, but in this case
symmetry breaking will not occur consistently.  The couplings $16 \cdot 16 \cdot \ov{126} + \ov{16} \cdot \ov{16} \cdot 126$
would provide the needed light--heavy mixing, if the SM singlet components of $16$ and $\ov{16}$ acquire VEVs.  Being the
light field, one should not allow $16 \cdot \ov{16}$ mass term, nor the coupling $16 \cdot \ov{16}. 210$.  Setting the
$F$--terms associated with the $16$ and $\ov{16}$ fields to zero, one finds $\left \langle 16 \right \rangle
\left \langle \ov{126} \right \rangle = 0$, and $\left \langle \ov{16} \right \rangle
\left \langle 126 \right \rangle = 0$. One solution to these equations is $\left\langle 16 \right\rangle = 0,
\left\langle \ov{16} \right\rangle = 0$, implying that the needed light--heavy mixing is not induced.  If the other solution
where $\left\langle 126 \right\rangle = 0, \left\langle \ov{126} \right\rangle = 0$ is chosen, $F$--term associated with the
$126$ and $\ov{126}$ would lead to $\left\langle 16 \right \rangle = 0, \left\langle \ov{16} \right\rangle = 0$, again leading
to vanishing heavy--light mixing.

In spite of the various constraints that need to be satisfied, we have found two additional ways of realizing the missing partner
mechanism with the use of $126+\ov{126}$ pair that do not lead to light exotics.

\noindent{\bf (ii) Heavy  \{\mbox{\boldmath{$126+\ov{126} +  45$}}\} + Light  \{\mbox{\boldmath{$ 10 +  120$}}\} model:}
\vspace*{0.1in}

In this model the heavy Higgs sector consists of $\{126+\ov{126}+45\}$.
Since the $45$--plet does not contain color triplets or $SU(2)_L$ doublets ($45 = 1(0) + 10(4) + \ov{10}(-4) + 24(0)$ under
$SU(5) \times U(1)$), the heavy sector will have two pairs of doublets
and three pairs of color triplet/anti-triplets, all arising from the $126+\ov{126}$.  The light sector is taken to be
$\{10+120\}$ which contains three pairs of doublets and three pairs of triplets/anti-triplets.
The relevant superpotential for doublet--triplet splitting is
\begin{equation}
\label{45}
W_{DT}^{(ii)} = 10 \cdot 126 \cdot 45^2 + 10 \cdot \ov{126} \cdot 45^2 + 120 \cdot 126 \cdot 45 + 120 \cdot \ov{126} \cdot 45~.
\label{W2}
\end{equation}
The $45^2$ in the first two terms of Eq. (\ref{45}) act effectively as a $210$--plet of model (i).  DT splitting would then work in analogy
with Eq. (\ref{210}), although
in the present case there is only a single $10$--plet light state (as opposed to two light $10$--plets in Eq. (\ref{210})).  This difference arises because, unlike the heavy $210$
which contains a pair of doublets and triplets, the $45$--plet employed in Eq. (\ref{45}) contains neither of these fields.
Although this new model utilizes dimension four terms in Eq. (\ref{45}) and in the superpotential for symmetry breaking
(terms such as $45^4$ are necessary for the $SO(10)$ symmetry to break down to the Standard Model symmetry), the particle content of the model is
relatively simple, making this model attractive.  The exotics from the $120$--plet all acquire super--heavy
masses by pairing with components of $126+\ov{126}$ or $45$.  When the SM singlet from the $24(0)$ of $45$ acquires a VEV, the last two terms
of Eq. (\ref{W2}) would generate the following mass terms.
\beqn
\left\langle({45},24(0))\right\rangle \cdot({126}, 10(-6))\cdot (120, (\ov{10}(6)),
~~~~\langle ({45},24(0)) \rangle \cdot(\ov{126}, \ov{10}(6))\cdot (120, 10(-6)),\nonumber\\
\langle ({45},24(0)) \rangle \cdot({126}, 45(2))\cdot (120, \ov{45}(-2)),
~~~~\langle ({45},24(0)) \rangle \cdot(\ov{126}, \ov{45}(-2))\cdot (120, 45(2))~.
\label{exoticmass2}
\eeqn
Additionally, when the VEVs of $1(-6)$ of the $126$ and the $1(6)$ of the $\ov{126}$ are inserted in Eq. (\ref{W2}),
$(120,10(-6)) \cdot (45, \ov{10}(-4)) \cdot \left\langle (\ov{126},1(10))\right\rangle$ and $(120,\ov{10}(6)) \cdot (45, {10}(4)) \cdot \left\langle ({126},1(-10))
\right\rangle$ will be induced, providing additional mass corrections to the exotics from the 120.  Thus we see that all fragments
from the 120, except for the MSSM Higgs doublet components, have become massive.

The fermion mass matrices that arise from the symmetric Yukawa couplings of the 10--plet and the antisymmetric Yukawa couplings
of the 120--plet do not lead to fully consistent charged fermion masses and mixings.  A simple solution to fix this problem, without upsetting
the DT mechanism is provided in Sec. 8.  Adding equal number of light and heavy fields to the spectrum of the present model does
not upset DT splitting mechanism, since the counting of doublets and triplets is unaltered.
If a pair of $126+\ov{126}$ fields are added to the heavy and the light spectrum, $H_u$ and
$H_d$ of the MSSM would have components from the $\ov{126}$, which would correct the wrong mass relations.

\vspace*{0.1in}
\noindent{\bf (iii) Heavy  \{\mbox{\boldmath{$126+\ov{126} $}}\} + Light  \{\mbox{\boldmath{$ 10 +  120$}}\} model:}
\vspace*{0.1in}

The last possibility for realizing the missing partner mechanism utilizing the $\{126+\ov{126}\}$ heavy fields is the one developed
in Ref. \cite{SO(10)Missingpartner}, which we briefly summarize.  Here one uses a $\{126+\ov{126}\}$ pair of heavy fields with the feature
that these fields do not acquire VEVs.
There are additional heavy Higgs fields belonging to $\{210+16+\ov{16}\}$ which acquire VEVs and break the $SO(10)$ gauge
symmetry down to $SU(3)_C \times SU(2)_L \times U(1)_Y$.  The doublets and color triplets from these  $\{210+16+\ov{16}\}$
fields do not mix with the doublets and triplets from the heavy $126+\ov{126}$ fields, nor with those from the light sector.
The light fields consist of
$\{10+120\}$, which contain a total of three doublet pairs and three color triplet/anti-triplet pairs.  When the light sector
mixes with the heavy $\{126+\ov{126}\}$ fields, and not with the heavy $\{210+16+\ov{16}\}$ fields responsible for $SO(10)$
symmetry breaking, all three color triplet/anti-triplet pairs become massive by pairing with the three color triplets/anti-triplets
from the $\{126+\ov{126}\}$. Only two of the doublet pairs from the $\{10+120\}$ become massive by mixing, since the $\{126+\ov{126}\}$ contain
two pairs of doublets. Thus one pair of doublets becomes light.  The relevant superpotential couplings are
\begin{equation}
W_{DT}^{(iii)} = 10 \cdot 126 \cdot 210 + 10 \cdot \ov{126} \cdot 210 + 120 \cdot 126 \cdot 210 + 120 \cdot \ov{126} \cdot 210~.
\end{equation}
Once the $75$ plet of $SU(5)$  in the $210$ of $SO(10)$ acquires a VEV,
all exotics from the 120 would acquire masses, as in Eq. (\ref{exoticmass}) of model (i).
The main difference of this model, compared to Eq. (\ref{210}) is that here the $\{126+\ov{126}\}$ heavy fields do not acquire VEVs,
and as a result the light doublet count is different.

It was shown in Ref. \cite{SO(10)Missingpartner} that there exists a $U(1)$ symmetry that
forbids all unwanted couplings to sufficiently high order.  This includes the
absence of mass terms for the light fields.  In this paper we simply use the supersymmetric non-renormalization
theorem to set the light field masses (or effective masses) to small values.

\section{Missing Partner Mechanism Anchored by $560+\ov{560}$}

As mentioned in the introduction, the second smallest irreducible representation after $126(\ov{126})$  with the feature
that it contains $\ov{50}(50)$ of $SU(5)$ needed for missing partner mechanism is $560 (\ov{560})$.  An additional nice feature of the
$560 (\ov{560})$ set is that it also contains the $75(\ov{75})$--plet representation.
Unlike the case of $126+\ov{126}$ where
a 210--plet was needed for this purpose, here there is no need for any other Higgs fields for symmetry breaking.  We shall
demonstrate how the $560+\ov{560}$ pair of Higgs breaks the $SO(10)$ symmetry down to $SU(3)_C \times SU(2)_L \times U(1)_Y$ in
one step in Sec. 6.  Here we study its attributes for realizing the missing partner mechanism in a simple and non-technical way,
with  more technical discussions given in Secs. 5 and 6.
The $560$ is a tensor--spinor representation, denoted as $\Psi_{\mu \nu}^\alpha$, where $\alpha=1-16$ is the spinor index, and
$\mu, \nu = 1-10$ are the tensor indices.  $\Psi_{\mu\nu}^\alpha$ obeys the following conditions: $\Gamma_\mu \Psi_{\mu \nu}^\alpha = 0$,
and $\Psi_{\mu \nu}^\alpha = - \Psi_{\nu \mu}^\alpha$, where $\Gamma_\mu$ are the $SO(10)$ gamma matrices.  With these conditions, $\Psi_{\mu\nu}^\alpha$ has 560 independent components.
The $560+\ov{560}$ multiplets will have to mix with a certain set of light fields.  To see the various possibilities, let us
note the decomposition of $560$ under $SU(5) \times U(1)$ \cite{slansky}:
\begin{eqnarray}
\label{560}
560 &=& 1(-5)+\overline{5}(3)+\ov{10}(-9)+10(-1)_1+10(-1)_2+24(-5)+40(-1)\nonumber \nonumber\\
&+&45(7)+\ov{45}(3)+\ov{50}(3)+\ov{70}(3)+75(-5)+175(-1)~.
\end{eqnarray}
The $560$ contains one up--type Higgs doublet, three down--type doublets, one color triplet and four color anti--triplets.
This follows from Eq. (\ref{560}), but can also be inferred from the decomposition of $560$ under $SU(2)_L \times SU(2)_R \times
SU(4)_C$ given in Ref. \cite{slansky}.  Thus $560+\ov{560}$ contain four pairs of Higgs doublets and five pairs of color
triplets/anti-triplets.  If these fields mix with light fields containing five pairs of doublets and triplets, one pair
of Higgs doublets from the light fields will remain light, to be identified as $H_u$ and $H_d$ of MSSM.

To see the various possibilities for the light fields, it is necessary to find out the decomposition of
$560 \times 560$, which is not readily available \cite{slansky}.  We have evaluated this from the publicly available
numerical program LiE \cite{lie}.  Since we employ only one pair of $560 + \ov{560}$, it is important to see which components
appear in the symmetric combination.  The product rule, in the symmetric and antisymmetric combinations, are given by
\begin{eqnarray}
(560 \times 560)_{s} &=& 10 + 126_1 + 126_2 + \ov{126} + 320 + 210' + 1728_1 + 1728_2 + 2970_1+2970_2 \nonumber \\
&+& 3696+4410+4950+\ov{4950} + 10560+ 6930' + 36750 + 27720 + 46800 ~, \nonumber \\
(560 \times 560)_{a} &=& 120_1 + 120_2 + 320 + 1728_1 + 1728_2 + 2970 + 3696_1 + 3696_2 \nonumber \\
&+& 4312_1 + 4312_2 + 10560 + 36750 + 34398 + 48114 ~.
\label{decomp}
\end{eqnarray}

\vspace*{0.1in}
\noindent{\bf (iv) Heavy  \{\mbox{\boldmath{$560+\ov{560} $}}\} + Light  \{\mbox{\boldmath{$  2 \times 10 +  320$}}\} model:}
\vspace*{0.1in}

From Eq. (\ref{decomp}), an essentially unique possibility for DT splitting via the missing partner mechanism emerges.
This uses $\{2 \times 10 + 320\}$--plets as the light fields.  The 320 contains
three pairs of doublets and triplets, and with the two pairs of doublets and triplets from the two 10--plets, this choice 
becomes consistent with pairing up all five color triplets, while pairing
only four sets of doublets with those from the $560+\ov{560}$.  The 320 multiplet is interesting and to our knowledge
has not been utilized in model building in particle physics.
The 320--plet can be taken to be a three index tensor which is anti-symmmetric in the first two
indices with the totally
anti-symmetric part (120) taken out and is traceless so that 450-120-10=320.  Under $SU(5) \times U(1)$ it decomposes as
\beqn
320= 5(2) + \bar 5(-2) + 40(-6) + \ov{40}(6) + 45(2) +\ov{45}(-2) + 70(2) + \ov{70}(-2),
\label{320}
\eeqn
which shows that it contains 3 doublet and 3 triplet/anti-triplet pairs.

We assume  a superpotential of the form
\beqn
{\mathsf W^{\prime}}=  M_{560}
~560\cdot\ov{560} + 560\cdot560\cdot 320 + \ov{560}\cdot\ov{560}\cdot 320 +
560\cdot560\cdot10_i + \ov{560}\cdot\ov{560}\cdot10_i
\label{Wp}
\eeqn
where the index $i=1,2$.  No mass terms are included for the $10$-plets and the $320$--plet.
We note that the heavy sector, i.e., $560+\ov{560}$,  gives us
4 doublets pairs and 5 triplet\a pairs (which enter in the DT splitting mechanism)
while the light sector consisting of $2\times 10 + 320$ Higgs gives us
2D+2T from $2\times 10$ and  3D+3T from 320 so that one has a total of 5D+5T from the light  sector.
Thus when the heavy and light sectors  mix we are left with just one pair of light Higgs doublets
while all other components of  $560+\ov{560}$ and of $2\times 10 + 320$ will be super--heavy.

Now we show how all exotics from the 320 become massive in this model.  
The light 320 has the following set of exotics: $\overline{40}(6)$, $45(2)$, $70(2)$, ${40}(-6)$, $\ov{45}(-2), \ov{70}(-2)$.
The exotics will combine with the heavy fields in $560+\ov{560}$ via the couplings of Eq. (\ref{Wp}).
Mass growth for the exotics occurs  as follows.
  \beqn
   <({560},R_v,-5)>\cdot({560}, {40}, -1)\cdot (320, \ov{40}, 6),
~~~~<(\ov{560},\ov{R}_v,5)>\cdot(\ov{560}, \ov{40}, 1)\cdot (320, 40, -6), \nonumber\\
<(560,R_v,-5)>\cdot(560, \ov{70}, 3)\cdot (320, 70, 2),
~~~~ <(\ov{560},\ov{R}_v,5)>\cdot(\ov{560}, 70, -3)\cdot (320, \ov{70}, -2),\nonumber\\
<({560},R_v,-5)>\cdot({560}, {45}, 7)\cdot (320, \ov{45}, -2),
~~~~<(\ov{560},\ov{R}_v,5)>\cdot(\ov{560}, \ov{45}, -7)\cdot (320, 45, 2), \nonumber\\
<({560},R_v,-5)>\cdot({560}, \ov{45}, 3)\cdot (320, {45}, 2),
~~~~<(\ov{560},\ov{R}_v,5)>\cdot(\ov{560}, {45}, -3)\cdot (320, \ov{45}, -2),
\eeqn
where $R_v =1, 24, 75$ are the $SU(5)$ representations from the $560+\ov{560}$ that develop VEVs.
We see that there is a one to one pairing of the light exotics with the heavy ones which will
make all the light exotics heavy. The remaining
components are just the  $5$--plets and $\bar 5$--plets in Eq. (\ref{320})
which enter in the  DT splitting realizing the missing partner
mechanism. So we find that the only massless  Higgs fields left after spontaneous breaking are  a pair of Higgs doublets and all the rest  become super--heavy and are removed from the low energy spectrum.

One might wonder if one of the two 10--plets in the light sector can be replaced by a $16+\ov{16}$, which also contain
one pair of doublets and one triplets.  Although one can write higher dimensional terms of the type $560\cdot560\cdot 16^2$
for mixing color triplets, the $16$ must develop a VEV along its SM singlet direction for this purpose.  Setting the
$F$--term for $16$ and $560$ would show that one of the two VEVs should vanish, in which case the missing partner mechanism
is not effective.

We note in passing that each of these models
have  $SU(5)$ couplings of type  $10\cdot10\cdot5$ and  $\overline{10}\cdot\overline{10}\cdot\overline{5}$.
Thus a flipped symmetry breaking chain, i.e.,
$SO(10)\to SU(5)_F\times U(1) \to SU(3)_C\times SU(2)_L\times U(1)_Y$ appears possible.

The results of the analysis, both with $126+\ov{126}$ and $560+\ov{560}$ heavy fields are summarized in Table 1.

{{\scriptsize
\begin{table}[htb]
\begin{center}
\begin{tabular}{|c|c|c|c|c|c|}
\hline
\hline
&&&&&\\
 Model &  Heavy Fields & Light Fields &Pairs of D and T & Pairs of D and T  &  Residual Set \\
 &   &   & in Heavy Fields   & in Light Fields  & of Light Modes   \\
 &&&&&\\
   \hline
   \hline
   &&&&&\\
 (i) &  $126+\overline{126}+210$&$ 2\times10+120$&(2D+3T)+(D+T)&(2D+2T)+(2D+2T)& 1D\\
 &&&&&\\
   \hline
   &&&&&\\
   (ii) &  $126+\overline{126}+45$&$ 10+120$&(2D+3T)&(D+T)+(2D+2T)& 1D\\
 &&&&&\\
   \hline
   &&&&&\\
   (iii) &  $126+\overline{126}$&$ 10+120$&(2D+3T)&(D+T)+(2D+2T)& 1D\\
 &&&&&\\
   \hline
   &&&&&\\
(iv)  & $560+\overline{560}$&$1\times 320 + 2 \times 10$&4D+5T& (3D+3T)+ (2D+2T) &1D \\
&&&&&\\
\hline
\hline
\end{tabular}\\~\\
\label{t1}
\caption{
Exhibition of Higgs doublet pairs (D) consisting of up--type and down--type Higgs doublets,
and Higgs triplet/anti-triplet (T) pairs in the $SO(10)$ missing partner models discussed in this work.
In case (iii) the $126+\ov{126}$ heavy fields do not acquire super--heavy VEVs. Additional $210+16+\ov{16}$
fields, which do not mix with the $126 + \ov{126}$, are utilized for $SO(10)$ symmetry breaking in this case.}
\end{center}
\end{table}
}}

\section{$\mathbf{560}$ and $\overline{\mathbf{560}}$ as Constrained $\mathbf{2^{\textnormal{\bf nd}}}$ Rank Antisymmetric Tensor-Spinor Multiplets}

In this section we build up the technology to deal with the $560+\ov{560}$ multiplets quantitatively.  These results will
be applied in Sec. 6 to study the one step symmetry breaking of $SO(10)$ down to the SM,
 and in Sec. 7 for the computation of the doublet and the triplet  Higgs mass matrices using $320+2\times 10$
 light Higgs sector.
As mentioned in Sec. 4, for the analysis to follow, it is convenient to consider an $SU(5)\times U(1)_X$
decomposition of the $560$ multiplet. Thus the 560 multiplet under $SU(5)\times U(1)_X$ is
\begin{eqnarray}\label{SU(5)decomposition}
   560&=&1(-5)[{\bf U}]~+~\overline{5}(3)[{\bf U}_i]~+~\overline{10}(-9)[{\bf U}_{ij}]~+~10(-1)[{ {\bf U}}^{ij}]~+~10(-1)[{\widehat {\bf U}}^{ij}]~+~24(-5)[{\bf U}^i_j]\nonumber\\
   &&+~40(-1)[{\bf U}^{ijk}_l]~+ ~45(7)[{\bf U}^{ij}_k]~+~\overline{45}(3)[{\bf U}_{ij}^k]~+~\overline{50}(3)[{\bf U}_{ijk}^{lm}]~+~\overline{70}(3)[{\bf U}_{(S)ij}^k]\nonumber\\
   &&+~75(-5)[{\bf U}_{kl}^{ij}]~+~175(-1)
     [{\bf U}_{[ijk]l}^m],
     \label{tensorstructure}
\end{eqnarray}
     where, for example, $1(-5)$ means that it is a singlet of $SU(5)$ with a $U(1)$ quantum number
     of $-5$ and
     where quantities in the brackets represent the tensorial structure of each $SU(5)$ multiplet.
     The 560 multiplet of $SO(10)$ is a  16-plet spinor with two tensor indices and  we represent this
     tensor-spinor by  $|\Theta_{\mu\nu}^{560}>$,  where $|\Theta_{\mu\nu}^{560}>$
is anti-symmetric in the indices $\mu\nu$ and
  satisfies the constraint $\Gamma_{\mu} |\Theta_{\mu\nu}^{560}> =0$.
Here $\mu,\nu$ are the $SO(10)$ indices, i.e., $\mu,\nu =1,2,\cdots 10$, and
$\Gamma_{\mu}$ are the $SO(10)$ gamma matrices which satisfy the Clifford algebra  relation
$\{\Gamma_{\mu}, \Gamma_{\nu}\} = 2\delta_{\mu\nu} I$.
  The irreducible tensor-spinor $|\Theta_{\mu\nu}^{560}>$ can be obtained from the reducible
  tensor-spinor  $|\Theta_{\mu\nu}^{720}>$  by subtraction of the 160 plet given by
  $\Gamma_{\mu} |\Theta_{\mu\nu}^{720}>$.

To proceed further, a useful approach is in terms of an
oscillator expansion where one defines $\Gamma_{\mu}$ in terms of $SU(5)$  oscillators as follows\cite{ms,wilczek}:
 $\Gamma_{2i}=(b_i+ b_i^{\dagger}), \Gamma_{2i-1}= -i(b_i-b_i^{\dagger}), i=1,\cdots,5$ where
 $b_i, b_i^{\dagger}$ satisfy the algebra $\{b_i,b_j^{\dagger}\}=\delta_{ij}$,
$\{b_i,b_j\}=0$ and $\{b_i^{\dagger}, b_j^{\dagger}\} = 0$.  In terms of $b_i, b_i^{\dagger}$ the
  720 (=16$\times$45) multiplet has the $SU(5)$ oscillator expansion:
$|\Theta_{\mu\nu}^{720}>=|0>{\theta}_{\mu\nu}+\frac{1}{2}b_i^{\dagger}b_j^{\dagger}|0>{\theta}_{\mu\nu}^{ij}+\frac{1}{24}\epsilon^{ijklm}
     b_j^{\dagger}b_k^{\dagger}b_l^{\dagger}b_m^{\dagger}|0>{\theta}_{i\mu\nu}$,
where as stated earlier $\mu,\nu,....$ are $SO(10)$ indices, while $i,j,k,l,...$ and $SU(5)$ indices.

In the decomposition of $SO(10)$ tensors into $SU(5)$ tensors it is convenient to first
express the $SO(10)$ tensors in terms of a specific set of $SU(5)$ reducible tensors. 
 These can then be further decomposed into irreducible $SU(5)$ tensors. The explicit computations of the
 $SO(10)$ couplings involving the $560$ multiplet  require in addition the techniques of the
 Basic Theorem developed in \cite{ns}  where one expresses the interaction structure using
 a Wick expansion in terms of the $SU(5)$ oscillators.

We now explicitly state the expansion of the 560 dimensional tensor-spinor in its $SU(5)$
oscillator modes
where we have implemented the constraint $\Gamma_{\mu} |\Theta_{\mu\nu}^{560}> =0$.
The result of a detailed analysis gives

\begin{eqnarray}
|\Theta_{c_xc_y}^{560}>&=&|0>{\bf U}^{xy}+\frac{1}{2}b_i^{\dagger}b_j^{\dagger}|0>
\left[\epsilon^{ijklm}{\bf U}^{xy}_{klm}\right.\nonumber\\
&&\left.+\frac{1}{96}\left(\epsilon^{ijkly}{\bf U}^{x}_{kl}
-\epsilon^{ijklx}{\bf U}^{y}_{kl}\right)-\frac{1}{288}\epsilon^{ijkxy}{\bf U}_k\right]+\frac{1}{24}\epsilon^{ijklm}
     b_j^{\dagger}b_k^{\dagger}b_l^{\dagger}b_m^{\dagger}|0>{\bf U}^{xy}_i,\label{tensor1}\\
     \nonumber\\
     |\Theta_{\bar{c}_x\bar{c}_y}^{560}>&=&|0>{\bf U}_{xy}+\frac{1}{2}b_i^{\dagger}b_j^{\dagger}|0>
\left[{\bf U}^{ij}_{xy}+\frac{1}{3}\left(\delta^i_y{\bf U}^j_x-\delta^i_x{\bf U}^j_y+\delta^j_x{\bf U}^i_y-\delta^j_y{\bf U}^i_x\right)\right.\nonumber\\
&&+\left.\frac{1}{20}\left(\delta^i_y\delta^j_x-\delta^i_x\delta^j_y\right){\bf U}\right]+\frac{1}{24}\epsilon^{ijklm}
     b_j^{\dagger}b_k^{\dagger}b_l^{\dagger}b_m^{\dagger}|0>\left[\epsilon_{nopxy}{\bf U}^{nop}_i\right.\nonumber\\
     &&\left. +\epsilon_{inoxy}\widehat{{\bf U}}^{no}\right],\label{tensor2}\\
     \nonumber\\
     |\Theta_{{c}_x\bar{c}_y}^{560}>&=&|0>\left[{\bf U}^x_y+\frac{1}{5}\delta^{x}_y{\bf U}\right]+\frac{1}{2}b_i^{\dagger}b_j^{\dagger}|0>
\left[\frac{1}{4}\left(\delta^i_y{\bf U}^{jx}-\delta^j_y{\bf U}^{ix}\right)\right.\nonumber\\
&&-\left.\frac{1}{72}\left(4\delta^x_y\widehat{{\bf U}}^{ij}
-\delta^j_y\widehat{{\bf U}}^{ix}+\delta^i_y\widehat{{\bf U}}^{jx}\right)-\frac{1}{6}{\bf U}^{ijx}_y+\epsilon^{ijklm}{\bf U}^{x}_{[klm]y}\right]\nonumber\\
&&+\frac{1}{24}\epsilon^{ijklm}
     b_j^{\dagger}b_k^{\dagger}b_l^{\dagger}b_m^{\dagger}|0>\left[\frac{1}{24}\left(5\delta^x_y{\bf U}_i-\delta^x_i{\bf U}_y\right)
     -\frac{1}{2}\left({\bf U}^{x}_{iy}+{\bf U}^{x}_{(S)iy}\right)
     \right].\label{tensor3}
\end{eqnarray}

\begin{eqnarray}
 |\overline{\Theta}_{{c}_x{c}_y}^{\overline{560}}>&=&b_1^{\dagger}b_2^{\dagger}b_3^{\dagger}b_4^{\dagger}b_5^{\dagger}|0>{\bf \overline{U}}^{xy}+\frac{1}{12}\epsilon^{ijklm}b_k^{\dagger}b_l^{\dagger}b_m^{\dagger}|0>
\left[{\bf \overline{U}}_{ij}^{xy}+\frac{1}{3}\left(\delta_i^y{\bf \overline{U}}_j^x-\delta_i^x{\bf \overline{U}}_j^y+\delta_j^x{\bf \overline{U}}_i^y\right.\right.\nonumber\\
&&\left.\left.-\delta_j^y{\bf \overline{U}}_i^x\right)+\frac{1}{20}\left(\delta_i^y\delta_j^x-\delta_i^x\delta_j^y\right){\bf \overline{U}}\right]+
    b_i^{\dagger}|0>\left[\epsilon^{nopxy}{\bf \overline{U}}_{nop}^i+\epsilon^{inoxy}\widehat{{\bf \overline{U}}}_{no}\right],\label{tensor4}\\
     \nonumber\\
|\overline{\Theta}_{\bar{c}_x\bar{c}_y}^{\overline{560}}>&=&b_1^{\dagger}b_2^{\dagger}b_3^{\dagger}b_4^{\dagger}b_5^{\dagger}|0>{\bf \overline{U}}_{xy}+\frac{1}{12}\epsilon^{ijklm}b_k^{\dagger}b_l^{\dagger}b_m^{\dagger}|0>
\left[\epsilon_{ijpqr}{\bf \overline{U}}_{xy}^{pqr}\right.\nonumber\\
&&\left.+\frac{1}{96}\left(\epsilon_{ijpqy}{\bf \overline{U}}_{x}^{pq}
-\epsilon_{ijpqx}{\bf \overline{U}}_{y}^{pq}\right)-\frac{1}{288}\epsilon_{ijpxy}{\bf \overline{U}}^p\right]+b_i^{\dagger}|0>{\bf \overline{U}}_{xy}^i,\label{tensor5}\\
     \nonumber\\
     |\overline{\Theta}_{{c}_x\bar{c}_y}^{\overline{560}}>&=&b_1^{\dagger}b_2^{\dagger}b_3^{\dagger}b_4^{\dagger}b_5^{\dagger}|0>\left[{\bf \overline{U}}^x_y+\frac{1}{5}\delta^{x}_y{\bf \overline{U}}\right]+\frac{1}{12}\epsilon^{ijklm}b_k^{\dagger}b_l^{\dagger}b_m^{\dagger}|0>
\left[\frac{1}{4}\left(\delta^x_i{\bf \overline{U}}_{jy}-\delta_j^x{\bf \overline{U}}_{iy}\right)\right.\nonumber\\
&&-\left.\frac{1}{72}\left(4\delta^x_y\widehat{{\bf \overline{U}}}_{ij}
-\delta_j^x\widehat{{\bf \overline{U}}}_{iy}+\delta_i^x\widehat{{\bf \overline{U}}}_{jy}\right)-\frac{1}{6}{\bf \overline{U}}_{ijy}^x+\epsilon_{ijpqr}{\bf \overline{U}}_{y}^{[pqr]x}\right]\nonumber\\
&&+b_i^{\dagger}|0>\left[\frac{1}{24}\left(5\delta^x_y{\bf \overline{U}}^i-\delta_y^i{\bf \overline{U}}^x\right)
     -\frac{1}{2}\left({\bf \overline{U}}_{y}^{ix}+{\bf \overline{U}}_{(S)y}^{ix}\right)
     \right].\label{tensor6}
\end{eqnarray}
Eqs. (\ref{tensor1}-\ref{tensor6}) show how the various  $SU(5)$  components given in Eq. (\ref{tensorstructure}) enter in the oscillator decomposition of the
constrained 560 multiplet.

\section{A More Unified $SO(10)$:  A One Scale Breaking of $SO(10)$ GUT Symmetry with $560+\ov{560}$}

In this section we give an analysis of the breaking of the GUT symmetry  when the  $560$ multiplet develops a VEV.
As mentioned in Sec. 1  we have three possible candidates for the VEV
  formation in the 560 multiplet. In $SU(5)\times U(1)_X$ decomposition they are the
 elements  $1(-5),~ 24(-5),~ 75(-5)$ in the $560$ and similarly $\bar 1(5),~ \overline{24}(5),~ \overline{75}(5)$
 in $\overline{560}$.  Suppose one breaks the GUT symmetry using just the $75$
 multiplet.  In this case there will be  a  large number of unwanted pseudo-Goldstone bosons
left over. To avoid this situation we will consider the case where
$1(-5),~ 24(-5),~ 75(-5)$ and their counterparts all develop VEVs. Indeed
one finds that the VEV growth for all the three fields is automatic unless one does an
extreme fine tuning.
Thus, in the analysis given below we  allow all the three multiplets to develop VEVs and
look for spontaneous breaking for this general case. (See also the Appendix).

In the analysis of spontaneous  breaking we will use
a combination of a bilinear term $560.\ov{560}$ and a  quartic term $(560.\ov{560})^2$ to develop VEVs.
 The bilinear term necessary for the GUT symmetry breaking is simply the mass term for the 560/$\overline{560}$ multiplet.
To generate the necessary quartic coupling involving $560$ and $\overline{560}$ multiplets, we start with the simplest possible contraction of $560$ and $\overline{560}$ with the $45$-plet of $SO(10)$, $\Phi$.  The relevant superpotential then takes the form
\begin{eqnarray}
\mathsf{W}=\frac{1}{2!}M_{45}\Phi_{\mu\nu}^{45}\Phi_{\mu\nu}^{45}+\lambda_{45}<\Theta_{\mu\nu}^{560*}|B|
\overline{\Theta}_{\nu\sigma}^{\overline{560}}>\Phi_{\mu\sigma}^{45}+M_{560}<\Theta_{\mu\nu}^{560*}|B|
\overline{\Theta}_{\mu\nu}^{\overline{560}}>.\label{GUTsymmetry}
\end{eqnarray}
The first term in $\mathsf{W}$ represents the mass term for the 45-dimensional anti-symmetric tensor and
the $B$ entering Eq. (\ref{GUTsymmetry}) is the $SO(10)$ charge conjugation operator so that
 $\displaystyle B=-\imath\prod_{i=1}^{5}\left(b_i-b_i^{\dagger}\right)$.
 Integrating out the 45-plet and giving the following VEVs to the  1-plets, the  24-plets, and the 75-plets
of $SU(5)$ contained in the 560 and $\overline{560}$,
\begin{eqnarray}
{<{\bf  U}>\choose <{\bf  \overline{U}}>}\equiv{{\bf  S}_{(1)}\choose {\bf  \overline{S}}_{(1)}},~~~~&&
{<{\bf  U}_{\beta}^{\alpha}>\choose <{\bf  \overline{U}}_{\beta}^{\alpha}>}=\frac{1}{3}\delta^{\alpha}_{\beta}{{\bf  S}_{(24)}\choose {\bf  \overline{S}}_{(24)}},\nonumber\\
{<{\bf  U}_{b}^{a}>\choose <{\bf  \overline{U}}_{b}^{a}>}=-\frac{1}{2}\delta^{a}_{b}{{\bf  S}_{(24)}\choose {\bf  \overline{S}}_{(24)}},~~~~&&
{<{\bf  U}_{\gamma\sigma}^{\alpha\beta}>\choose <{\bf  \overline{U}}_{\gamma\sigma}^{\alpha\beta}>}=\frac{1}{6}\left(\delta^{\alpha}_{\gamma}\delta^{\beta}_{\sigma}-
\delta^{\alpha}_{\sigma}\delta^{\beta}_{\gamma}\right){{\bf  S}_{(75)}\choose {\bf  \overline{S}}_{(75)}}\nonumber\\
{<{\bf  U}_{cd}^{ab}>\choose <{\bf  \overline{U}}_{cd}^{ab}>}=\frac{1}{2}\left(\delta^{a}_{c}\delta^{b}_{d}-
\delta^{a}_{d}\delta^{b}_{c}\right){{\bf  S}_{(75)}\choose {\bf  \overline{S}}_{(75)}},~~~~&&
{<{\bf  U}_{\beta b}^{\alpha a}>\choose <{\bf  \overline{U}}_{\beta b}^{\alpha a}>}=-\frac{1}{6}\delta^{a}_{b}\delta^{\alpha}_{\beta}{{\bf  S}_{(75)}\choose {\bf  \overline{S}}_{(75)}},
\end{eqnarray}
we get,
\begin{eqnarray}
\mathsf{W}
&=&\frac{\lambda_{45}^2}{4M_{45}}\left[-\left(\frac{7}{27}\right){\bf  S}_{(75)}^2{\bf  \overline{S}}_{(75)}^2
-\left(\frac{10}{81}\right){\bf  S}_{(75)}^2{\bf  \overline{S}}_{(75)}{\bf  \overline{S}}_{(24)}-\left(\frac{10}{81}\right){\bf  S}_{(75)}{\bf  \overline{S}}_{(75)}^2{\bf{S}}_{(24)}\right.\nonumber\\
&&\left.-\left(\frac{349}{972}\right){\bf  S}_{(75)}{\bf  \overline{S}}_{(75)}{\bf  S}_{(24)}{\bf  \overline{S}}_{(24)}
+\left(\frac{7}{720}\right){\bf  S}_{(75)}{\bf  \overline{S}}_{(75)}{\bf  S}_{(24)}{\bf  \overline{S}}_{(1)}+\left(\frac{7}{720}\right){\bf  S}_{(75)}{\bf  \overline{S}}_{(75)}{\bf  \overline{S}}_{(24)}{\bf  {S}}_{(1)}\right.\nonumber\\
&&\left.-\left(\frac{1}{50}\right){\bf  S}_{(75)}{\bf  \overline{S}}_{(75)}{\bf  S}_{(1)}{\bf  \overline{S}}_{(1)}
-\left(\frac{125}{1944}\right){\bf  \overline{S}}_{(75)}^2{\bf {S}}_{(24)}^2-\left(\frac{125}{1944}\right){\bf  S}_{(75)}^2{\bf  \overline{S}}_{(24)}^2\right.\nonumber\\
&&\left.+\left(\frac{5}{216}\right){\bf  S}_{(75)}{\bf  \overline{S}}_{(24)}{\bf  S}_{(24)}{\bf  \overline{S}}_{(1)}
+\left(\frac{5}{216}\right){\bf  S}_{(75)}{\bf  \overline{S}}_{(24)}^2{\bf  S}_{(1)}+\left(\frac{5}{216}\right){\bf  \overline{S}}_{(75)}{\bf  {S}}_{(24)}^2{\bf  \overline{S}}_{(1)}\right.\nonumber\\
&&\left.+\left(\frac{5}{216}\right){\bf  \overline{S}}_{(75)}{\bf  S}_{(24)}{\bf  \overline{S}}_{(24)}{\bf  S}_{(1)}
-\left(\frac{73}{2916}\right){\bf  S}_{(75)}{\bf  \overline{S}}_{(24)}^2{\bf  S}_{(24)}-\left(\frac{73}{2916}\right){\bf  \overline{S}}_{(75)}{\bf  {S}}_{(24)}^2{\bf  \overline{S}}_{(24)}\right.\nonumber\\
&&\left.+\left(\frac{1015}{17496}\right){\bf  \overline{S}}_{(24)}^2{\bf  S}_{(24)}^2
+\left(\frac{1}{648}\right){\bf  S}_{(24)}^2{\bf  \overline{S}}_{(24)}{\bf  \overline{S}}_{(1)}+\left(\frac{1}{648}\right){\bf  \overline{S}}_{(24)}^2{\bf  {S}}_{(24)}{\bf {S}}_{(1)}\right.\nonumber\\
&&\left.-\left(\frac{1}{480}\right){\bf {S}}_{(24)}^2{\bf  \overline{S}}_{(1)}^2
-\left(\frac{1}{480}\right){\bf  \overline{S}}_{(24)}^2{\bf {S}}_{(1)}^2+\left(\frac{1}{144}\right){\bf  \overline{S}}_{(24)}{\bf  {S}}_{(24)}{\bf {S}}_{(1)}{\bf  \overline{S}}_{(1)}\right.\nonumber\\
&&\left.+\left(\frac{1}{500}\right){\bf  \overline{S}}_{(1)}^2{\bf  S}_{(1)}^2\right]+\imath M_{560}\left[-{\bf  \overline{S}}_{(75)}
{\bf {S}}_{(75)}+\left(\frac{10}{9}\right){\bf  \overline{S}}_{(24)}
{\bf {S}}_{(24)}+\left(\frac{7}{20}\right){\bf  \overline{S}}_{(1)}
{\bf {S}}_{(1)}\right].\label{GUTsymmetryvevs}
\end{eqnarray}
Here, for example, the 24--plet of $SU(5)$ contained in the 560--plet of $SO(10)$ has the Standard Model singlet denoted by ${\bf S}_{(24)}$, while, the 75--plet of $SU(5)$ contained in the $\overline{560}$--plet of $SO(10)$ has the Standard Model singlet denoted by ${\bf \overline{S}}_{(75)}$, etc..

\begin{table}[t]
\begin{center}
\begin{tabular}{|c|c|c|c|}
\multicolumn{4}{c}{Table 2: ${\bf {S}}_{(1)}={\bf  \overline{S}}_{(1)}$, ${\bf {S}}_{(24)}={\bf  \overline{S}}_{(24)}$, ${\bf {S}}_{(75)}={\bf  \overline{S}}_{(75)}$}\cr \hline \hline

 $M_{45}\cdot M_{560}~({\rm GeV}^2)$&${\bf {S}}_{(1)}~({\rm GeV})$& ${\bf {S}}_{(24)}~({\rm GeV})$&
 ${\bf  {S}}_{(75)}~({\rm GeV})$
 \\
 \hline\hline

\multirow{2}{*}{$10^{30}$}& {\small{$(-1.2+\imath 1.4)\times10^{16}$}} & {\small{$(27+\imath 5.7)\times 10^{14}$}}&
{\small{$(6.4+\imath 25)\times 10^{14}$}}\\
\cline{2-4}

&{\small{$(-5.4-\imath 5.4)\times10^{12}$}} & {\small{$(-2.5+\imath 2.5)\times 10^{14}$}}&
{\small{$(2-\imath 2)\times 10^{15}$}} \\
\hline\hline

\multirow{2}{*}{$10^{31}$}& {\small{$(4.4-\imath 3.8)\times10^{16}$}} & {\small{$(1.8+\imath 8.4)\times 10^{15}$}}&
{\small{$(7.8+\imath 2)\times 10^{15}$}}\\
\cline{2-4}

&{\small{$(-1.7+\imath 1.7)\times10^{13}$}} & {\small{$(-7.9+\imath 7.9)\times 10^{14}$}}&
{\small{$(6.5-\imath 6.5)\times 10^{15}$}} \\
\hline\hline

\multirow{2}{*}{$10^{32}$}& {\small{$(1.2-\imath 1.4)\times10^{17}$}} & {\small{$(-27-\imath 5.7)\times 10^{15}$}}&
{\small{$(-6.4-\imath 25)\times 10^{15}$}}\\
\cline{2-4}

&{\small{$(-5.4+\imath 5.4)\times10^{13}$}} & {\small{$(-2.5+\imath 2.5)\times 10^{15}$}}&
{\small{$(2-\imath 2)\times 10^{16}$}} \\
\hline\hline

\multirow{2}{*}{$10^{33}$}& {\small{$(-3.8+\imath 4.4)\times10^{17}$}} & {\small{$(8.4+\imath 1.8)\times 10^{16}$}}&
{\small{$(2+\imath 7.9)\times 10^{16}$}}\\
\cline{2-4}

&{\small{$(1.7-\imath 1.7)\times10^{14}$}} & {\small{$(7.9-\imath 7.9)\times 10^{15}$}}&
{\small{$(-6.5+\imath 6.5)\times 10^{16}$}} \\
\hline\hline

\multirow{2}{*}{$10^{34}$}& {\small{$(1.4-\imath 1.2)\times10^{18}$}} & {\small{$(5.7+\imath 27)\times 10^{16}$}}&
{\small{$(25+\imath 6.4)\times 10^{16}$}}\\
\cline{2-4}

&{\small{$(-5.4+\imath 5.4)\times10^{14}$}} & {\small{$(-2.5+\imath 2.5)\times 10^{16}$}}&
{\small{$(2-\imath 2)\times 10^{17}$}} \\
\hline
\hline
\end{tabular}
\label{t4}
\caption{Numerical estimates of the VEVs of the singlet, the 24--plet and the 75--plet fields in the $560+\ov{560}$ multiplets
in the spontaneous breaking of the $SO(10)$ gauge symmetry at the  GUT scale.}
\end{center}
\end{table}

Using Eq. (\ref{GUTsymmetryvevs}) we now look for solutions for which
$$\frac{\partial \mathsf{W}}{\partial{\bf  S}_{(75)}}=0,~~~\frac{\partial \mathsf{W}}{\partial{\bf  \overline{S}}_{(75)}}=0,~~~\frac{\partial \mathsf{W}}{\partial{\bf  S}_{(24)}}=0,~~~\frac{\partial \mathsf{W}}{\partial{\bf  \overline{S}}_{(24)}}=0,~~~\frac{\partial \mathsf{W}}{\partial{\bf  S}_{(1)}}=0,~~~\frac{\partial \mathsf{W}}{\partial{\bf  \overline{S}}_{(1)}}=0,$$
are satisfied simultaneously.
In general there are six VEVs to be determined. Since these are coupled cubic equations in six parameters
it is not possible to solve these equations analytically. Further, the number of allowed solutions is very large.
Thus the  solutions to the VEV equations were carried out numerically on Mathematica.
Since the number of allowed solutions is rather large
 we display here only a few sample cases. Thus, for the case when  the VEVs of $1(-5)$ and $\bar 1(5)$ are
 taken  equal,
 and similarly for    $24(-5)$ and $\ov{24}(5)$, and for  $75(-5)$ and $\ov{75}(5)$,  we display in Table 4  a set  of 10 solutions, two for each  value of $M_{45}\cdot M_{560}$.
We display one such solution below
\beqn
 & M_{45}\cdot M_{560}=10^{32} ~{\rm GeV}^2,&\non
&{\bf  S}_{(1)}= (1.2-\imath 1.4)\times10^{17},~
{\bf  S}_{(24)}=
  {\tiny{(-2.7-\imath .57)\times 10^{16}}},~
{\bf  S}_{(75)}={\tiny{(-.64-\imath 2.5)\times 10^{16}}},&  ~~~~
\label{vevs}
\eeqn
where the VEVs are all in the unit of GeV. The VEVs of Eq. (\ref{vevs}) are close to the conventional
unification scale of $2\times 10^{16}$ GeV. We note that in determining the VEVs only the product
of $M_{45}\cdot M_{560}$ enters.\\

\section{Higgs Doublets and Higgs Triplet Mass Matrices in $560+\ov{560}+ (2\times 10+ 320)$ Model}
\begin{table}[t]
\centering
{\bf } \vspace{.3cm}
\vspace{.3cm}
\begin{center}
\begin{tabular}{|c||c|c|c|c|c|c|c|c|c|c|c|c|}

\hline
 & & & & & & & & & & & &\\
   & $5_{_{320}}$& $\overline{5}_{_{320}}$ &$40_{_{320}}$& $\overline{40}_{_{320}}$& $45_{_{320}}$ & $\overline{45}_{_{320}}$ & $70_{_{320}}$ & $\overline{70}_{_{320}}$
& $5_{1_{{_{10}}_{_1}}}$ & $\overline{5}_{1_{{_{10}}_{_1}}}$ & $5_{2_{{_{10}}_{_2}}}$ & $\overline{5}_{2_{{_{10}}_{_2}}}$ \\
\hline
\hline
 & & & & & & & & & & & &\\
$\overline{5}_{_{560}}$     &   $\mathsf d_1$ &     &&      &     $\mathsf d_{10}$       &             &              &                 &      $\mathsf d_{24}$       &               & $\mathsf d_{30}$            &        \\
\hline
 & & & & & & & & & & & &\\
$40_{_{560}}$ & & & & $\mathsf d_{9}$& & & & & & & &\\
\hline
 & & & & & & & & & & & &\\
$\overline{45}_{_{560}}$      &   $\mathsf d_2$     &     &&     &     $\mathsf d_{11}$       &             &              $\mathsf d_{18}$ &                  &       $\mathsf d_{25}$      &               &    $\mathsf d_{31}$           &        \\
\hline
 & & & & & & & & & & & &\\
$45_{_{560}}$ &        &     &&      &            &      $\mathsf d_{14}$       &              &      $\mathsf d_{21}$            &              &          $\mathsf d_{27}$     &               &   $\mathsf d_{33}$     \\
\hline
 & & & & & & & & & & & &\\
$\overline{70}_{_{560}}$    & $\mathsf{d_3}$       &  &&         &       $\mathsf d_{12}$     &            &       $\mathsf d_{19}$       &                  &              &               &               &        \\
\hline
 & & & & & & & & & & & &\\
 $5_{_{\overline{560}}}$   &        &  $\mathsf d_5$ &&        &            &     $\mathsf d_{15}$         &            &                  &              &         $\mathsf d_{28}$      &               &     $\mathsf d_{34}$   \\
\hline
 & & & & & & & & & & & &\\
$\overline{40}_{_{\overline{560}}}$& & & $\mathsf d_{8}$& & & & & & & & &\\
\hline
 & & & & & & & & & & & &\\
   $45_{_{\overline{560}}}$    &        &   $\mathsf d_6$ &&        &            &  $\mathsf d_{16}$           &              &      $\mathsf d_{22}$            &              &      $\mathsf d_{29}$         &               &  $\mathsf d_{35}$     \\
\hline
 & & & & & & & & & & & &\\
  $\overline{45}_{_{\overline{560}}}$     &  $\mathsf d_4$      &    &&       &   $\mathsf d_{13}$         &             &      $\mathsf d_{20}$        &                &   $\mathsf d_{26}$           &               &     $\mathsf d_{32}$          &        \\
\hline
 & & & & & & & & & & & &\\
  $70_{_{\overline{560}}}$    &        & $\mathsf d_7$    &&      &            &       $\mathsf d_{17}$      &              &       $\mathsf d_{23}$           &              &               &               &        \\
\hline
\end{tabular}
\label{t5}
\caption{A list of non-vanishing mass terms  for the Higgs doublets in the $560+\ov{560} +320+2\times 10$ missing partner  model.}
\end{center}
 \end{table}

In Table 3 we exhibit in detail the mass generation  for the Higgs  doublets.
Here
the entries $\mathsf d_{1}-\mathsf d_{19}$ arise from the mixings of the $560$ multiplet with 320 while the entries $\mathsf d_{20}-\mathsf d_{31}$ arise from mixings
with  the Higgs 10--plets $10_1,~ 10_2$.   The entries $\mathsf d_{1}-\mathsf d_{31}$  are listed below.
\begin{eqnarray}
\mathsf d_{1}=\left(\matrix{<1_{_{560}}(-5)>\cr
<24_{_{560}}(-5)>}\right)\cdot \overline{5}_{_{560}}(3)\cdot5_{_{320}}(2), ~&&~
\mathsf d_{2}=\left(\matrix{<24_{_{560}}(-5)>\cr
<75_{_{560}}(-5)>}\right)\cdot \overline{45}_{_{560}}(3)\cdot5_{_{320}}(2)\nonumber\\
\mathsf d_{3}=\left(<24_{_{560}}(-5)>\right)\cdot \overline{70}_{_{560}}(3)\cdot{5}_{_{320}}(2), ~&&~
\mathsf d_{4}=\left(\matrix{<\overline{24}_{_{\overline{560}}}(5)>\cr <\overline{75}_{_{\overline{560}}}(5)>}\right)\cdot \overline{45}_{_{\overline{560}}}(-7)\cdot5_{_{320}}(2), \nonumber\\
\mathsf d_{5}=\left(\matrix{<\overline{1}_{_{\overline{560}}}(5)>\cr<\overline{24}_{_{\overline{560}}}(5)>}\right)\cdot 5_{_{\overline{560}}}(-3)\cdot\overline{5}_{_{320}}(-2)~&&~
\mathsf d_{6}=\left(\matrix{<\overline{24}_{_{\overline{560}}}(5)>\cr<\overline{75}_{_{\overline{560}}}(5)>}\right)\cdot 45_{_{\overline{560}}}(-3)\cdot\overline{5}_{_{320}}(-2)\nonumber\\
\mathsf d_{7}=\left(<\overline{24}_{_{\overline{560}}}(5)>\right)\cdot 70_{_{\overline{560}}}(-3)\cdot\overline{5}_{_{320}}(-2),~&&~
\mathsf d_{8}=\left(\matrix{<\overline{1}_{_{\overline{560}}}(5)>\cr<\overline{24}_{_{\overline{560}}}(5)>\cr
<\overline{75}_{_{\overline{560}}}(5)>}\right)\cdot \overline{40}_{_{\overline{560}}}(1)\cdot40_{_{320}}(-6)\nonumber\\
\mathsf d_{9}=\left(\matrix{<1_{_{560}}(-5)>\cr<24_{_{560}}(-5)>\cr
<75_{_{560}}(-5)>}\right)\cdot 40_{_{{560}}}(-1)\cdot\overline{40}_{_{320}}(6), ~&&~
\mathsf d_{10}=\left(\matrix{<24_{_{560}}(-5)>\cr
<75_{_{560}}(-5)>}\right)\cdot \overline{5}_{_{560}}(3)\cdot45_{_{320}}(2)\nonumber\\
\mathsf d_{11}=\left(\matrix{<1_{_{560}}(-5)>\cr<24_{_{560}}(-5)>\cr
<75_{_{560}}(-5)>}\right)\cdot \overline{45}_{_{560}}(3)\cdot 45_{_{320}}(2), ~&&~
\mathsf d_{12}=\left(\matrix{<24_{_{560}}(-5)>\cr
<75_{_{560}}(-5)>}\right)\cdot \overline{70}_{_{560}}(3)\cdot 45_{_{320}}(2)\nonumber\\
\mathsf d_{13}=\left(\matrix{<\overline{1}_{_{\overline{560}}}(5)>\cr<\overline{24}_{_{\overline{560}}}(5)>\cr
<\overline{75}_{_{\overline{560}}}(5)>}\right)\cdot \overline{45}_{_{\overline{560}}}(-7)\cdot 45_{_{320}}(2), ~&&~
\mathsf d_{14}=\left(\matrix{<1_{_{560}}(-5)>\cr<24_{_{560}}(-5)>\cr
<75_{_{560}}(-5)>}\right)\cdot 45_{_{560}}(7)\cdot \overline{45}_{_{320}}(-2)\nonumber\\
\mathsf d_{15}=\left(\matrix{<\overline{24}_{_{\overline{560}}}(5)>\cr<\overline{75}_{_{\overline{560}}}(5)>}\right)\cdot 5_{_{\overline{560}}}(-3)\cdot \overline{45}_{_{320}}(-2), ~&&~
\mathsf d_{16}=\left(\matrix{<\overline{1}_{_{\overline{560}}}(5)>\cr<\overline{24}_{_{\overline{560}}}(5)>\cr
<\overline{75}_{_{\overline{560}}}(5)>}\right)\cdot 45_{_{\overline{560}}}(-3)\cdot \overline{45}_{_{320}}(-2)\nonumber\\
\mathsf d_{17}=\left(\matrix{<\overline{24}_{_{\overline{560}}}(5)>\cr<\overline{75}_{_{\overline{560}}}(5)>}\right)\cdot 70_{_{\overline{560}}}(-3)\cdot \overline{45}_{_{320}}(-2), ~&&~
\mathsf d_{18}=\left(\matrix{<24_{_{560}}(-5)>\cr
<75_{_{560}}(-5)>}\right)\cdot \overline{45}_{_{560}}(3)\cdot 70_{_{320}}(2)\nonumber\\
\mathsf d_{19}=\left(\matrix{<1_{_{560}}(-5)>\cr<24_{_{560}}(-5)>\cr
<75_{_{560}}(-5)>}\right)\cdot \overline{70}_{_{560}}(3)\cdot 70_{_{320}}(2), ~&&~
\mathsf d_{20}=\left(\matrix{<\overline{24}_{_{\overline{560}}}(5)>\cr<\overline{75}_{_{\overline{560}}}(5)>}\right)\cdot \overline{45}_{_{\overline{560}}}(-7)\cdot 70_{_{320}}(2)\nonumber\\
\mathsf d_{21}=\left(\matrix{<24_{_{560}}(-5)>\cr
<75_{_{560}}(-5)>}\right)\cdot 45_{_{560}}(7)\cdot 70_{_{320}}(-2),~&&~\mathsf d_{22}=\left(\matrix{<\overline{24}_{_{\overline{560}}}(5)>\cr
<\overline{75}_{_{\overline{560}}}(5)>}\right)\cdot 45_{_{\overline{560}}}(-3)\cdot \overline{70}_{_{320}}(-2)\nonumber\\
\mathsf d_{23}=\left(\matrix{<\overline{1}_{_{\overline{560}}}(5)>\cr<\overline{24}_{_{\overline{560}}}(5)>\cr
<\overline{75}_{_{\overline{560}}}(5)>}\right)\cdot 70_{_{\overline{560}}}(-3)\cdot \overline{70}_{_{320}}(-2), ~&&~
\mathsf d_{24}=\left(\matrix{<1_{_{560}}(-5)>\cr
<24_{_{560}}(-5)>}\right)\cdot \overline{5}_{_{560}}(3)\cdot5_{1_{{_{10}}_{_1}}}(2)\nonumber\\
\mathsf d_{25}=\left(\matrix{<24_{_{560}}(-5)>\cr
<75_{_{560}}(-5)>}\right)\cdot \overline{45}_{_{560}}(3)\cdot5_{1_{{_{10}}_{_1}}}(2), ~&&~
\mathsf d_{26}=\left(\matrix{<\overline{24}_{_{\overline{560}}}(5)>\cr<\overline{75}_{_{\overline{560}}}(5)>}\right)\cdot \overline{45}_{_{\overline{560}}}(-7)\cdot 5_{1_{{_{10}}_{_1}}}(2)\nonumber\\
\mathsf d_{27}=\left(\matrix{<24_{_{560}}(-5)>\cr
<75_{_{560}}(-5)>}\right)\cdot 45_{_{560}}(7)\cdot \overline{5}_{1_{{_{10}}_{_1}}}(-2), ~&&~
\mathsf d_{28}=\left(\matrix{<\overline{1}_{_{\overline{560}}}(5)>\cr<\overline{24}_{_{\overline{560}}}(5)>}\right)\cdot 5_{_{\overline{560}}}(-3)\cdot \overline{5}_{1_{{_{10}}_{_1}}}(-2)\nonumber\\
\mathsf d_{29}=\left(\matrix{<\overline{24}_{_{\overline{560}}}(5)>\cr
<\overline{75}_{_{\overline{560}}}(5)>}\right)\cdot 45_{_{\overline{560}}}(-3)\cdot \overline{5}_{1_{{_{10}}_{_1}}}(-2), ~&&~
\mathsf d_{30}=\left(\matrix{<1_{_{560}}(-5)>\cr
<24_{_{560}}(-5)>}\right)\cdot \overline{5}_{_{560}}(3)\cdot 5_{2_{{_{10}}_{_2}}}(2)\nonumber\\
\mathsf d_{31}=\left(\matrix{<24_{_{560}}(-5)>\cr
<75_{_{560}}(-5)>}\right)\cdot \overline{45}_{_{560}}(3)\cdot 5_{2_{{_{10}}_{_2}}}(2), ~&&~
\mathsf d_{32}=\left(\matrix{<\overline{24}_{_{\overline{560}}}(5)>\cr
<\overline{75}_{_{\overline{560}}}(5)>}\right)\cdot \overline{45}_{_{\overline{560}}}(-7)\cdot 5_{2_{{_{10}}_{_2}}}(2)\nonumber\\
\mathsf d_{33}=\left(\matrix{<24_{_{560}}(-5)>\cr
<75_{_{560}}(-5)>}\right)\cdot 45_{_{560}}(7)\cdot \overline{5}_{2_{{_{10}}_{_2}}}(-2), ~&&~
\mathsf d_{34}=\left(\matrix{<\overline{1}_{_{\overline{560}}}(5)>\cr<\overline{24}_{_{\overline{560}}}(5)>}\right)\cdot 5_{_{\overline{560}}}(-3)\cdot\overline{5}_{2_{{_{10}}_{_2}}}(-2)\nonumber\\
\mathsf d_{35}=\left(\matrix{<\overline{24}_{_{\overline{560}}}(5)>\cr
<\overline{75}_{_{\overline{560}}}(5)>}\right)\cdot 45_{_{\overline{560}}}(-3)\cdot\overline{5}_{2_{{_{10}}_{_2}}}(-2). ~&&~ \nonumber\\
\end{eqnarray}
The mass matrix arising from  Table 3 leads to one pair of massless Higgs doublets while all the
other Higgs doublets become heavy.

The triplet/anti-triplet mass matrix is given in Table 4.
Here $\mathsf t_{1}-\mathsf t_{35}$ are defined similar to  $\mathsf d_{1}-\mathsf d_{31}$,  except that one is extracting the Higgs triplets/anti-triplets couplings rather than the
Higgs doublet couplings. The new contributions to the Higgs triplet\a mass matrix arise from the terms $\mathsf T_{1}- \mathsf T_{8}$ which are given by \begin{eqnarray}
\mathsf T_{1}=<75_{_{560}}(-5)>\cdot \overline{50}_{_{560}}(3)\cdot 5_{_{320}}(2), ~&&~
\mathsf T_{2}=<\overline{75}_{_{\overline{560}}}(5)\cdot 50_{_{\overline{560}}}(-3)\cdot\overline{5}_{_{320}}(-2)\nonumber\\
\mathsf T_{3}=\left(\matrix{<24_{_{560}}(-5)>\cr
<75_{_{560}}(-5)>}\right)\cdot \overline{50}_{_{560}}(3)\cdot 45_{_{320}}(2), ~&&~
\mathsf T_{4}=\left(\matrix{<\overline{24}_{_{\overline{560}}}(5)>\cr<\overline{75}_{_{\overline{560}}}(5)>}\right)\cdot 50_{_{\overline{560}}}(-3)\cdot\overline{45}_{_{320}}(-2)\nonumber\\
\mathsf T_{5}=
<75_{_{560}}(-5)>\cdot \overline{50}_{_{560}}(3)\cdot 5_{1_{{_{10}}_{_1}}}(2), ~&&~
\mathsf T_{6}=<\overline{75}_{_{\overline{560}}}(5)\cdot 50_{_{\overline{560}}}(-3)\cdot\overline{5}_{1_{{_{10}}_{_1}}}(-2)\nonumber\\
\mathsf T_{7}=<75_{_{560}}(-5)>\cdot \overline{50}_{_{560}}(3)\cdot 5_{2_{{_{10}}_{_2}}}(2), ~&&~
\mathsf T_{8}=<\overline{75}_{_{\overline{560}}}(5)\cdot 50_{_{\overline{560}}}(-3)\cdot\overline{5}_{2_{{_{10}}_{_2}}}(-2).\nonumber
\end{eqnarray}
The  triplet/anti-triplet mass matrix arising from the entries
of Table 4 is rank 10 and thus all the Higgs triplets/anti-triplets become super heavy and are removed from the low energy spectrum of the theory.  Further,  as shown in Sec. 4 all the remaining components of the 320 multiplet become massive.
Thus at the end we are left with only one light pair of $SU(2)_L$ Higgs doublets
and all the remaining Higgs fields will be super--heavy.
\begin{table}[t]
\centering
{\bf } \vspace{.3cm}
\vspace{.3cm}
\begin{center}
\begin{tabular}{|c||c|c|c|c|c|c|c|c|c|c|c|c|}

\hline
 & & & & & & & & & & & &\\
   & $5_{_{320}}$& $\overline{5}_{_{320}}$ &$40_{_{320}}$& $\overline{40}_{_{320}}$& $45_{_{320}}$ & $\overline{45}_{_{320}}$ & $70_{_{320}}$ & $\overline{70}_{_{320}}$
& $5_{1_{{_{10}}_{_1}}}$ & $\overline{5}_{1_{{_{10}}_{_1}}}$ & $5_{2_{{_{10}}_{_2}}}$ & $\overline{5}_{2_{{_{10}}_{_2}}}$ \\
\hline
\hline
 & & & & & & & & & & & &\\
$\overline{5}_{_{560}}$     &   $\mathsf t_1$ &     &&      &     $\mathsf t_{10}$       &             &              &                 &      $\mathsf t_{24}$       &               & $\mathsf t_{30}$            &        \\
\hline
 & & & & & & & & & & & &\\
$40_{_{560}}$ & & & & $\mathsf t_{9}$& & & & & & & &\\
\hline
 & & & & & & & & & & & &\\
$\overline{45}_{_{560}}$      &   $\mathsf t_2$     &     &&     &     $\mathsf t_{11}$       &             &              $\mathsf t_{18}$ &                  &       $\mathsf t_{25}$      &               &    $\mathsf t_{31}$           &        \\
\hline
 & & & & & & & & & & & &\\
$45_{_{560}}$ &        &     &&      &            &      $\mathsf t_{14}$       &              &      $\mathsf t_{21}$            &              &          $\mathsf t_{27}$     &               &   $\mathsf t_{33}$     \\
\hline
 & & & & & & & & & & & &\\
$\overline{70}_{_{560}}$    & $\mathsf{t_3}$       &  &&         &       $\mathsf t_{12}$     &            &       $\mathsf t_{19}$       &                  &              &               &               &        \\
\hline
 & & & & & & & & & & & &\\
 $5_{_{\overline{560}}}$   &        &  $\mathsf t_5$ &&        &            &     $\mathsf t_{15}$         &            &                  &              &         $\mathsf t_{28}$      &               &     $\mathsf t_{34}$   \\
\hline
 & & & & & & & & & & & &\\
$\overline{40}_{_{\overline{560}}}$& & & $\mathsf t_{8}$& & & & & & & & &\\
\hline
 & & & & & & & & & & & &\\
   $45_{_{\overline{560}}}$    &        &   $\mathsf t_6$ &&        &            &  $\mathsf t_{16}$           &              &      $\mathsf t_{22}$            &              &      $\mathsf t_{29}$         &               &  $\mathsf t_{35}$     \\
\hline
 & & & & & & & & & & & &\\
  $\overline{45}_{_{\overline{560}}}$     &  $\mathsf t_4$      &    &&       &   $\mathsf t_{13}$         &             &      $\mathsf t_{20}$        &                &   $\mathsf t_{26}$           &               &     $\mathsf t_{32}$          &        \\
\hline
 & & & & & & & & & & & &\\
  $70_{_{\overline{560}}}$    &        & $\mathsf t_7$    &&      &            &       $\mathsf t_{17}$      &              &       $\mathsf t_{23}$           &              &               &               &        \\
\hline
 & & & & & & & & & & & &\\
$\overline{50}_{_{560}}$    &    $\mathsf T_{1}$     &    &&       &    $\mathsf T_{3}$       &             &              &           &   $\mathsf T_{5}$           &        &     $\mathsf T_{7}$           &        \\
  \hline
   & & & & & & & & & & & &\\
  $50_{_{\overline{560}}}$    &        &    $\mathsf T_{2}$       &   &&         &      $\mathsf T_{4}$      &              &       $$           &              &     $\mathsf T_{6}$          &               &   $\mathsf T_{8}$      \\
\hline
\end{tabular}
\label{t6}
\caption{A list of non-vanishing mass  terms for the Higgs triplets/anti-triplets  in the $560+\ov{560} +320+2\times 10$
missing partner model.}
\end{center}
 \end{table}

\section{Fermion Masses, $b-t-\tau$ Unification and Proton Decay}
The matter fields as usual will reside in three
generations of 16--plets and these will have cubic couplings with the light Higgs doublets.
Typically in models discussed in Table 1 the light Higgs doublets will be linear combination
of doublets in $10$--plets and in $120$--plets of Higgs  fields. In addition one can produce
non-minimal models  by adding  an arbitrary (equal) number of light and heavy generations of $10$'s
$120$'s and $126+\ov{126}$'s.
As an example,  one may add
 a light and a heavy $126'+ \overline{126'}$ with a mixing between the
two terms, i.e., one adds additional non-minimal terms as follows
\beqn
W_{\rm non-min}= M \, 126_h'.\overline{126'}_h + M' \,(126'_h.\overline{126'}_l+126'_l.\overline{126'}_h).
\label{yuk4}
\eeqn
Here $126_h'$ is the heavy Higgs multiplets  with mass $M$ while $126'_l$ is the light Higgs multiplet
which has no mass term but it mixes with the heavy multiplet $126'$. In this case the light
Higgs will in general  be a linear combination of $10+120+ 126$.
The addition of these combinations of  light and heavy fields
will not disturb the overall count of the light pair of doublets.
Using the above technique  fully realistic models for fermion masses
are possible in all cases discussed in Table 1.

Specifically, consider models (ii) and (iii) listed in Table 1.  The MSSM Higgs fields $(H_u,\, H_d$) have components
in a 10 and a 120 in these cases.  If we add to this model a heavy and light $126+\ov{126}$ as shown in Eq. (\ref{yuk4}),
$(H_u,\,H_d$) will also have components from the light $126+\ov{126}$. Such models can generate fully consistent quark
and lepton masses, and also large neutrino mixing angles.  The case where the Yukawa couplings to the light 120 is
small would be a special case of a minimal $SO(10)$ model that has been widely studied \cite{bm}.  This model actually predicts
a relatively large value of $\sin^2 2\theta_{13} \approx 0.085$, which is compatible with the recent results from T2K \cite{T2K}
and MINOS \cite{MINOS} experiments.  While the models presented here do not predict the value of $\theta_{13}$, this special case
suggests that they can certainly accommodate such large values for this neutrino mixing angle.  

Next we consider briefly the fermion masses and mixings in the 560+ $\overline{56} +
10_1+ 10_2+320$  model.  Since the 320 does not couple to $16_i$ matter multiplets, 
the symmetric coupling matrices of $10_i$ ($i=1,2$) would lead to the GUT scale degeneracy
of down--type quarks and charged leptons.  One could utilize the light--heavy $126+\ov{126}$ pair
as in Eq. (\ref{yuk4}) to correct these relations.  The heavy $\ov{126}$ can be used to generate large Majorana 
masss for the right--handed neutrinos.  Alternatively, one could use the non-renormalizable coupling
 $16_i\,\, 16_j \,\, \ov{560} \,\, \ov{560}$ couplings. With two 10--plets and a $\ov{126}$--plet of light fields,
 there are three symmetric Yukawa coupling matrices, giving enough freedom to correct all the wrong mass relations
 that would arise in the absence of the $\ov{126}$--plet.  Thus one can fit all the observables and
specifically one has the possibility of producing a large $\theta_{13}$ in the neutrino mixing
matrix.

We further note that since
  $H_u, H_d$ are  now linear combinations of several fields,
the couplings of $H_u$ and $H_d$ are not necessarily
equal at the unification scale. The prediction of $b-\tau$ unification still holds since
$b$ and $\tau$ reside in the same $SU(5)$ multiplet and their masses arise from the same
 $H_d$.
 However,  the usual
$SO(10)$ prediction that $b-t-\tau$ unification which requires large $\tan\beta$ \cite{Ananthanarayan:1991xp}
 need not hold in this model for the reasons given above, i.e., that
 the couplings of $H_u$ and of $H_d$ are not necessarily equal at the unification
scale.  Thus in such models it is possible to achieve a $b-t-\tau$ unification without the necessity of a
large $\tan\beta$.
Further, the Higgs triplet couplings in such models will as usual produce baryon and lepton number violating dimension five operators (For a review see \cite{Nath:2006ut}).
However, since the Higgs triplets/anti-triplets are linear combinations of several mass eigenstates
 it appears possible to suppress them via the cancellation mechanism \cite{Nath:2007eg}, i.e.,
 via  an appropriate choice of the mixing angles that enter in the linear combinations.
 Detailed analyses of these issues are outside the scope of this work.

\section{Conclusion}

 Models based on $SO(10)$ are desirable for a variety of reasons and continue to attract attention
 (For  recents work see \cite{Girrbach:2011an} and the references therein).
 In this paper we have analyzed the missing partner mechanism anchored by
 the $126+\ov{126}$ Higgs fields and found three consistent models.  
 We have also proposed and developed a new class of missing partner $SO(10)$ models anchored
 by the $560+\ov{560}$ multiplets.  Here, for
  the light states an essentially unique possibility is found,
 consisting of $\{2 \times 10 + 320\}$ multiplets.
This class of models allow for spontaneous breaking of the $SO(10)$
gauge group to the Standard Model gauge group in one step
at the unification scale, in contrast to
 the conventional $SO(10)$ models where one needs two scales, i.e., one scale to break the rank
 of the group,  and the second scale to reduce the gauge symmetry down to
 $SU(3)_C\times SU(2)_L\times U(1)_Y$.
   Further, the models lead to a natural doublet-triplet splitting via the missing partner mechanism
   where all exotic fields become heavy leaving  only a pair of light Higgs doublets.
The couplings of light Higgs doublets $H_d$ and $H_u$ are not necessarily equal at the unification
scale since the light Higgses are  linear combination of various multiplets.  Because of this the low
energy phenomenology of this  model is  different from a class of
$SO(10)$ models where the MSSM parameter $\tan\beta$ takes a large value approximately equal to $m_t/m_b$.
It is found that fully realistic models emerge for some cases,
while others  can be made realistic by addition of vector--like representations.
It appears possible to suppress proton decay via dimension five operators but
this analysis along with other phenomenological issues requires a further investigation.

We note that one of the important issues concerns the gauge coupling unification scale $M_G$.
As in well known gauge coupling unification is affected by heavy thresholds. A possibility of
cancelations in the threshold effects among different fragments of $560 +\overline{560}$ exists. However, this
requires a detailed computation of all the heavy thresholds in the theory which is outside the
scope of this work.

Finally, we note that the theory is not asymptotically free above the unification scale $M_G$ because
of the large number of degrees of freedom. However, the physics beyond the unification scale
is largely unknown because one is entering the domain where gravity becomes strong. This is specifically
the case when one has a large number of degrees of freedom N since as pointed out recently  \cite{bdv}  the
effective fundamental scale here is reduced by a factor $\sqrt {N}$ which thus lies close to the scale $M_G$. The above  implies that non-renormalizable interactions due to the
closeness of the fundamental scale  could be large and
must be included above $M_G$, which would redefine
the theory above this scale. More specifically, above the scale $M_G$ we should only
work with the UV complete theory in this case.

\noindent
{\bf  Acknowledgements:}
This research is supported in part by DOE grants
DE-FG02-04ER41306 and DE-FG02-ER46140 (KSB);
DOE Grant No. DE-FG02-91ER40626 (IG), and by NSF grants   PHY-0757959 and PHY-0969739 (PN).  We
wish to thank the Center for Theoretical Underground Physics and Related Areas (CETUP*) for hospitality
during this year's inaugural summer program in Lead, South Dakota, where part of this work was done.

  \section* {Appendix:  Further Details of spontaneous breaking with $560+\overline{560}$ Higgs
  and Absence of Goldstones}
  We discuss here some further aspects of spontaneous breaking with the $560+\overline{560}$
  of Higgs fields.
The simplest superpotential that would induce symmetry breaking $SO(10)\rightarrow SU(3)_C\times SU(2)_L\times U(1)_Y$ is of the form
\begin{eqnarray}
W=M~560\cdot\overline{560}+\frac{1}{M^{\prime}}(560\cdot\overline{560})_r\cdot(560\cdot\overline{560})_r
\label{A1}
\end{eqnarray}
where $r$ stands for the representation by which $560$ contracts.
 Let us  suppose the contraction is trivial, i.e., $r$ is a singlet of $SO(10)$.  In this case the Lagrangian
 will have an $SU(560)$ global symmetry and a VEV formation of any singlet of $560$ will lead to 559
 Goldstone bosons. To avoid this situation one must consider a non-trivial contraction and the simplest
 such contraction is when $r=45$, i.e., one considers the case
 $(560\cdot\overline{560})_{45}\cdot(560\cdot\overline{560})_{45}$.
   For the spontaneous symmetry breaking analysis it is useful
to keep the 45--plet in the spectrum while
analyzing the minimum of the potential, and then take the mass of the 45 to be very large
to remove it from the spectrum.  The superpotential couplings will then have  a mass  term for 45,
a  mass term $M$ for the 560, and in addition will have the following cubic coupling:
\begin{eqnarray}
560\cdot\overline{560}\cdot 45 &=& 1\cdot\overline{1}\cdot\widehat{1}+1\cdot\overline{24}\cdot\widehat{24}+24\cdot\overline{1}\cdot\widehat{24}+24\cdot\overline{24}\cdot\widehat{24}\nonumber
\\&+&24\cdot\overline{75}\cdot\widehat{24}+75\cdot\overline{24}\cdot\widehat{24}+75\cdot\overline{75}\cdot\widehat{24}
\label{A2}
\end{eqnarray}
Here the hatted components are from 45, the bars arise from the $\overline{560}$, and have opposite $U(1)_X$ charges compared to the un-barred components from $560$. Only the components which contain the $SM$ singlets
are shown  in Eq. (\ref{A2}). Now a VEV formation for the $SU(5)$ singlet alone will not break
the gauge symmetry completely down to the $SM$  gauge group and one would need in addition the VEV
formation for either the 24--plet or the 75--plet.  However, we will show that the VEV formation of either
the $24$--plet or the $75$ plet will lead to VEV formation for all the rest.

Let us consider the case when say the $24$ and $\overline{24}$--plets get VEVs. Then from the 4th term
on the right hand side of Eq. (\ref{A2}) one finds that $\widehat{24}$ will get a VEV since it is linear in
 $\widehat{24}$, and  then from the 5th and 6th terms on the right hand side of Eq. (\ref{A2}) one finds that
 $75$ and $\ov{75}$ will get VEVs. Further from the second and the third terms on the right hand side
 of Eq. (\ref{A2}) one finds that $1$ and $\bar 1$ will get VEVs.  Thus one arrives at the result that if
 24--plet gets a VEV, then $1, \bar 1, 75, \ov{75}$ all get VEVS.  A similar argument shows that if $75$ and
 $\overline{75}$ get VEVs then the rest, i.e., $1, \bar 1, 24, \ov{24}$ all get VEVs.
The above argument can be repeated if $r=210$ in Eq. (\ref{A1}).
Specifically the expansion of the cubic coupling $560\cdot\overline{560}\cdot 210$  gives the SM singlets as follows
 \begin{eqnarray}
560\cdot\overline{560}\cdot 210&=& 1\cdot\overline{1}\cdot\widehat{1}+24\cdot\overline{24}\cdot\widehat{1}+75\cdot\overline{75}\cdot\widehat{1}+1\cdot\overline{24}\cdot\widehat{24}+
24\cdot\overline{1}\cdot\widehat{24}\nonumber\\&&
+24\cdot\overline{24}\cdot\widehat{24}+24\cdot\overline{75}\cdot\widehat{24}
+
75\cdot\overline{24}\cdot\widehat{24}
+75\cdot\overline{75}\cdot\widehat{24}+1\cdot\overline{75}\cdot\widehat{75}+24\cdot\overline{24}\cdot\widehat{75}
\nonumber\\&&+
24\cdot\overline{75}\cdot\widehat{75}+75\cdot\overline{24}\cdot\widehat{75}+75\cdot\overline{75}\cdot\widehat{75}.
\label{A3}
\end{eqnarray}
where the hatted fields are from the 210. Eq. (\ref{A3}) is very similar  to
 Eq.(\ref{A2}) except that this time we also have $\widehat{75}$ in addition to $\widehat{1}$,
$\widehat{24}$.  However, the analysis in this case gives  exactly the same result, i.e.,
if either the $24, \ov{24}$ or the $75, \ov{75}$ get VEVs, then all the rest of the fields get VEVs as in the
$r=45$ case.

The quartic term in Eq. (\ref{A1})  can also arise by the contractions $(560\cdot{560})_{\ov{126}}$,  $(560\cdot{560})_{120}$ or $(560\cdot{560})_{10}$, which are all allowed. Among these, the 120 and 10 contractions will not generate quartic terms involving SM singlet, since these fields (10 and 120) do not contain SM singlets.  The 126 contraction is a possibility. However, in this case, the symmetry breaks down to $SU(5)$ only,  or else, many components of the 75-fragment of the 560 will remain massless. The superpotential in this case has mass terms for the 560 and for the 126, as well as two cubic terms:  $(560\cdot{560})_{\overline{126}}$  and $(\overline{560}\cdot\overline{560})_{{126}}$. The singlet components from the cubic terms are
\begin{eqnarray}
560\cdot 560\cdot \overline{126}&=& 1\cdot{1}\cdot\widehat{1}+24\cdot{24}\cdot\widehat{1}+75\cdot{75}\cdot\widehat{1}\nonumber\\
\overline{560}\cdot\overline{560}\cdot{126}&=& \overline{1}\cdot\overline{1}\cdot\widehat{\overline{1}}+\overline{24}\cdot\overline{24}\cdot\widehat{\overline{1}}
+\overline{75}\cdot\overline{75}\cdot\widehat{\overline{1}}
\label{A4}
\end{eqnarray}
where $\widehat{1}$ refers to the singlet fragment of $\overline{126}$ and $\widehat{\overline{1}}$ is the singlet fragment from the 126. The mass terms are explicitly  given by
\begin{eqnarray}
W_{mass}=M(1\cdot\overline{1}+24\cdot\overline{24}+75\cdot\overline{75})+M_{126}~\widehat{1}\cdot\widehat{\overline{1}}
\label{A5}
\end{eqnarray}

It is possible to obtain a minimum of Eqs. (\ref{A4}) and (\ref{A5}) with some of the VEVs zero. Thus, for example, one may have
$1=\overline{1}=24=\overline{24}=0$ while
 $\widehat{1}, \widehat{\overline{1}}, 75, \overline{75}$ all nonzero.
However, if such a solution is chosen, one would end up with many components of $75+\overline{75}$ being massless. The reason is that Eqs. (\ref{A4})-(\ref{A5}) have an $O(75)$ symmetry, with both the $75$ and $\overline{75}$ transforming as vectors. If the SM singlets in the 75 acquire VEVs, this global symmetry is broken down to $O(74)$, which generates many unwanted Goldstone bosons. On the other hand, if the singlets ($1, \overline{1}, \widehat{1}, \widehat{\overline{1}}$) from 560 and 126 are taken to have non-zero VEV's, there is no such problem with Goldstones. However, in this case the unbroken symmetry is $SU(5)$. This proves that integrating out 126 can only result in $SO(10)\rightarrow SU(5)$ symmetry breaking.
Next suppose we consider representations higher than 126 containing SM singlets that are in the contraction $(560\cdot 560)$.
 This contraction has odd number of $SO(10)$ vector indices (odd number from the two spinor indices, and even number of vector indices, resulting in odd number of indices). Odd number of vector indices cannot contain 24 or 75 fragments; only possible SM singlet is in the 1 of $SU(5)$. Then the argument given for the 126 contraction will go through for higher representations as well and
 one will have the $SO(10)\rightarrow SU(5)$ symmetry breaking. The above analysis implies that in the analysis of
 spontaneous breaking one must consider non-trivial contractions, i.e., with $r$ is Eq. (\ref{A1}) which is a non-singlet
 and specifically $45$ or $210$
 and further the VEV growths for all the relevant fields, i.e., $1, 24, 75$ in the 560 multiplet must be taken into account
 for consistency.


\begin{thebibliography}{999}

\bibitem{georgi}
H. Georgi, in Particles and Fields (edited by C.E. Carlson), A.I.P.,
1975; H. Fritzch and P. Minkowski, Ann. Phys. {\bf 93}, 193 (1975).

\bibitem{Babu:2005gx}
  K.~S.~Babu, I.~Gogoladze, P.~Nath and R.~M.~Syed,
  Phys.\ Rev.\  D {\bf 72}, 095011 (2005)
  [arXiv:hep-ph/0506312].

\bibitem{Babu:2006rp}
  K.~S.~Babu, I.~Gogoladze, P.~Nath and R.~M.~Syed,
  Phys.\ Rev.\  D {\bf 74}, 075004 (2006)
  [arXiv:hep-ph/0607244].

\bibitem{Nath:2005bx}
  P.~Nath and {R.~M.~Syed},
  JHEP {\bf 0602}, 022 (2006)
  [arXiv: hep-ph/0511172];
  Phys.\ Rev.\  D {\bf 81}, 037701 (2010)
  [arXiv:0909.2380 [hep-ph]].

\bibitem{DW}
S. Dimopoulos and F. Wilczek, Print-81-0600 (Santa Barbara);
K.~S.~Babu and S.~M.~Barr,
  Phys.\ Rev.\  D {\bf 48}, 5354 (1993).

\bibitem{SU(5)Missingpartner}
  A.~Masiero, D.~V.~Nanopoulos, K.~Tamvakis and T.~Yanagida,
  Phys.\ Lett.\  B {\bf 115}, 380 (1982);
  B.~Grinstein,
  Nucl.\ Phys.\  B {\bf 206}, 387 (1982).

\bibitem{SO(10)Missingpartner}
  K.~S.~Babu, I.~Gogoladze and Z.~Tavartkiladze,
  Phys.\ Lett.\  B {\bf 650}, 49 (2007)
  [arXiv:hep-ph/0612315].

  \bibitem{slansky}
 R.~Slansky,
  Phys.\ Rept.\  {\bf 79}, 1 (1981).

  \bibitem{aulakh}
   B.~Bajc, A.~Melfo, G.~Senjanovic and F.~Vissani,
    Phys.\ Rev.\  D {\bf 70}, 035007 (2004);
   T.~Fukuyama, A.~Ilakovac, T.~Kikuchi, S.~Meljanac and N.~Okada,
   J.\ Math.\ Phys.\  {\bf 46}, 033505 (2005);
  C.~S.~Aulakh, A.~Girdhar,
   Nucl.\ Phys.\  {\bf B711}, 275-313 (2005).

\bibitem{lie}

See: http://www-math.univ-poitiers.fr/~maavl/LiE/.

\bibitem{ms}
R.N. Mohapatra and B. Sakita, Phys. Rev. {\bf D21}, 1062 (1980).

\bibitem{wilczek}
F. Wilczek and A. Zee, Phys. Rev. {\bf D25}, 553 (1982).

\bibitem{ns}
 P.~Nath and {R.~M.~Syed},
Phys.\ Lett.\ B {\bf 506}, 68 (2001);
Nucl.\ Phys.\ B {\bf 618}, 138 (2001);
Nucl.\ Phys.\ B {\bf 676}, 64 (2004);
 R. M. Syed,
 arXiv: hep-ph/0411054;
   arXiv: hep-ph/0508153.
  

\bibitem{bm}
 K.~S.~Babu and R.~N.~Mohapatra, Phys.\ Rev.\ Lett.\  {\bf 70}, 2845 (1993);
 T.~Fukuyama and N.~Okada, JHEP {\bf 0211}, 011 (2002);
 B.~Bajc, G.~Senjanovic and F.~Vissani, Phys.\ Rev.\ Lett.\  {\bf 90}, 051802 (2003);
 C.~S.~Aulakh, B.~Bajc, A.~Melfo, G.~Senjanovic and F.~Vissani, Phys.\ Lett.\ B {\bf 588}, 196 (2004);
 B.~Dutta, Y.~Mimura and R.~N.~Mohapatra, Phys.\ Lett.\ B {\bf 603}, 35 (2004);
 K.~S.~Babu and C.~Macesanu, Phys.\ Rev.\ D {\bf 72}, 115003 (2005);
 S.~Bertolini, T.~Schwetz and M.~Malinsky, Phys.\ Rev.\ D {\bf 73}, 115012 (2006).

\bibitem{T2K}

 K.~Abe {\it et al.}  [T2K Collaboration],
 Phys.\ Rev.\ Lett.\  {\bf 107}, 041801 (2011).
 
 \bibitem{MINOS}
  P.~Adamson {\it et al.}  [MINOS Collaboration],
  Phys.\ Rev.\ Lett.\  {\bf 107}, 181802 (2011).



\bibitem{Ananthanarayan:1991xp}
  B.~Ananthanarayan, G.~Lazarides and Q.~Shafi,
  Phys.\ Rev.\  D {\bf 44}, 1613 (1991).

\bibitem{Nath:2006ut}
  P.~Nath and P.~Fileviez Perez,
  Phys.\ Rept.\  {\bf 441}, 191 (2007)
  [arXiv:hep-ph/0601023].

\bibitem{Nath:2007eg}
  P.~Nath and R.~M.~Syed,
  Phys.\ Rev.\  D {\bf 77}, 015015 (2008)
  [arXiv:0707.1332 [hep-ph]].

\bibitem{Girrbach:2011an}
  J.~Girrbach, S.~Jager, M.~Knopf, W.~Martens, U.~Nierste, C.~Scherrer and S.~Wiesenfeldt,
  arXiv:1101.6047 [hep-ph];
 L.~Di Luzio,
    [arXiv:1102.3590 [hep-ph]];
  V.~De Romeri, M.~Hirsch and M.~Malinsky,
  arXiv:1107.3412 [hep-ph].
  
  \bibitem{bdv}
 G.~Dvali,
  Fortsch.\ Phys.\ \ {\bf 58}, 528  (2010);
R. Brustein , G. Dvali, G. Veneziano,  JHEP 0910:085,2009.

\end{thebibliography}
\end{document}